\documentclass[10pt]{iopart}

\usepackage{iopams}
\usepackage{bm,mathrsfs,color}
\usepackage{graphicx}
\usepackage{braket}

\newcommand{\bv}{\bm{v}}
\newcommand{\bE}{\bm{E}}
\newcommand{\bB}{\bm{B}}
\newcommand{\bj}{\bm{j}}

\newcommand{\kp}{\bm{k}\cdot \bm{p}}
\newcommand{\hmu}{\hat{\mu}}
\newcommand{\hnu}{\hat{\nu}}
\newcommand{\hB}{\hat{B}}

\newcommand{\D}{\Delta}
\newcommand{\ve}{\varepsilon}

\begin{document}

\title[Magnetoresistance and valley degree of freedom in bulk bismuth]
{
Magnetoresistance and valley degree of freedom in bulk bismuth}

\author{Zengwei Zhu$^{1}$,
Beno\^{\i}t Fauqu\'e$^{2,3}$,
Kamran Behnia$^{2}$,
 \& Yuki Fuseya$^{2,4}$
 }

\address{
(1)Wuhan National High Magnetic Field Center and School of Physics, Huazhong University of Science and Technology,  Wuhan  430074, China\\
(2)Laboratoire Physique et Etude de Mat\'{e}riaux (CNRS-UPMC)ESPCI Paris, PSL Research University, 75005 Paris, France\\
(3)JEIP, USR 3573 CNRS, Coll\`ege de France, PSL Research University, 11, place Marcelin Berthelot, Paris Cedex 05 75231, France\\
(4)Department of Engineering Science, University of Electro-Communications, Chofu, Tokyo 182-8585, Japan}
\ead{zengwei.zhu@hust.edu.cn}
\ead{fuseya@uec.ac.jp}

\begin{abstract}
In this paper, we first review fundamental aspects of magnetoresistance in multi-valley systems based on the semiclassical theory. Then we will review experimental evidence and theoretical understanding of magnetoresistance in an archetypal multi-valley system, where the  electric conductivity is set by the sum of the contributions of different valleys. Bulk bismuth has three valleys with an extremely anisotropic effective mass. As a consequence the magnetoconductivity in each valley is extremely sensitive to the orientation of the magnetic field. Therefore, a rotating magnetic field plays the role of a valley valve tuning the contribution of each valley to the total conductivity. In addition to this simple semi-classical effect, other phenomena arise in the high-field limit as a consequence of an intricate Landau spectrum. In the vicinity of the quantum limit, the orientation of magnetic field  significantly affects the distribution of carriers in each valley, namely, the valley polarization is induced by the magnetic field. Moreover, experiment has found that well beyond the quantum limit, one or two valleys become totally empty. This is the only case in condensed-matter physics where a Fermi sea is completely dried up by a magnetic field without a metal-insulator transition. There have been two long-standing problems on bismuth near the quantum limit: the large anisotropic Zeeman splitting of holes, and the extra peaks in quantum oscillations, which cannot be assigned to any known Landau levels. These problems are solved by taking into account the interband effect due to the spin-orbit couplings for the former, and the contributions from the twinned crystal for the latter. Up to here, the whole spectrum can be interpreted within the one-particle theory. Finally, we will discuss transport and thermodynamic signatures of breaking of the valley symmetry in this system. By this term, we refer to the observed spontaneous loss of threefold symmetry at high magnetic field and low temperature. Its theoretical understanding is still missing. We will discuss possible explanations.
\end{abstract}

%
%
%
\maketitle
%
\ioptwocol
\tableofcontents

\section{Introduction}\label{Intro}
Magnetoresistance, the change in the electric resistance induced by the application of magnetic field, has two aspects of interests. One aspect is the magnetoresistance as the object of research itself from viewpoints of both basic and applied physics. The mechanisms of its large magnitude and its field dependence have been investigated for a long time since the age of Kapitza \cite{Kapitza1928}. The materials that exhibit large magnetoresistance have been used in various fields of application \cite{Daughton1999}. In the past few years, large and non-saturating magnetoresistance has attracted renewed interests following the recent observation in WTe$_2$ \cite{Ali2014}. The interests are rapidly expanding related to the unusual transport phenomena in topological materials of Dirac and Weyl fermion systems, such as in Cd$_3$As$_2$ \cite{TLiang2014} and NbP \cite{Shekhar2015}.

The other aspect is the magnetoresistance as the probe to measure the electronic properties in solids. The quantum oscillation in magnetoresistance, the so-called Shubnikov-de Haas effect, has been played a crucial role in solid state physics as a powerful tool to measure the Fermi surface together with its magnetic susceptibility version, the de Haas-van Alphen effect \cite{Shoenberg_book}. Besides this fermiology, the quantum oscillation measurements can also bring rich information on the spin-orbit coupling, which is one of the central issues in contemporary solid state physics, through the observation of the Zeeman splitting. The Zeeman splitting due to crystalline spin-orbit coupling is of increasing importance in a wide rage of fields, such as Berry phase physics \cite{Mikitik1999,DXiao2010,Murakawa2013}, spintronics \cite{Wolf2001,Zutic2004,Soumyanarayanan2016}, and nanophysics \cite{Murani2017}. Recently, it is also attracting an attention as the key for unraveling the longstanding mystery in heavy fermion systems, the hidden order in URu$_2$Si$_2$ \cite{Mydosh2011,Bastien2017}.

Surprisingly and interestingly, the origins of these two aspects --- large and non-saturating magnetoresistance and quantum oscillation in magnetoresistance --- are the two discoveries made almost at the same time \cite{Kapitza1928,Shubnikov1930}, and studying the same material, namely bismuth. This intriguing solid is the main subject of the present review.

In addition to these two historical aspects, there is another emergent aspect of magnetoresistance
 --- controlling the valley degree of freedom \cite{Rycerz2007,DXiao2007}. 
Many semiconductors (e.g., Si, Ge, AlAs, PbTe, transition-metal dichalcogenide), semimetals (e.g., Bi, Sb, As), and carbon based materials (diamond, graphene) have valley degree of freedom. While crystallographically equivalent valleys are degenerate at the ground state, they can be polarized by external fields, such as electric field \cite{Gunn1963,Butcher1967}, strain \cite{Gunawan2006,Gunawan2007} or circularly polarized light \cite{HZeng2012,Mak2012,Sallen2012}. The magnetic field can also generate such a valley polarized state, which is observed by the magneto-transport \cite{Shkolnikov2002,ZZhu2011,ZZhu2011b,ZZhu2012,Collaudin2015,ZZhu2017}, and the thermodynamic measurements \cite{LLi2008,Kuchler2014}.

Bismuth, which is one of the most studied crystals in solid state physics \cite{Dresselhaus1971,Edelman1976,Issi1979,Fuseya2015}, sits at the crossroads of the three trends: large and non-saturating magnetoresistance, Zeeman effect due to spin-orbit coupling, and the valley degree of freedom. As was mentioned, the magnetoresistance of bismuth has been investigated for a longtime since Kapitza's age, and it keeps increasing even up to 90.5 T \cite{ZZhu2017}. Although the main reason of the non-saturating behavior would be due to the perfect compensation between electron and hole carriers, the mechanism of its field dependence is still controversial \cite{Abrikosov1969,Abrikosov2003,Song2015,Owada2017}. The Zeeman splitting of electron carrier exhibit typical properties of Dirac electrons, whereas that of hole carrier was a half-a-century-old puzzle, which was recently solved as the interband effects of the spin-orbit coupling \cite{Fuseya2015b}.

\begin{figure}
	\begin{center}
		\includegraphics[width=5cm]{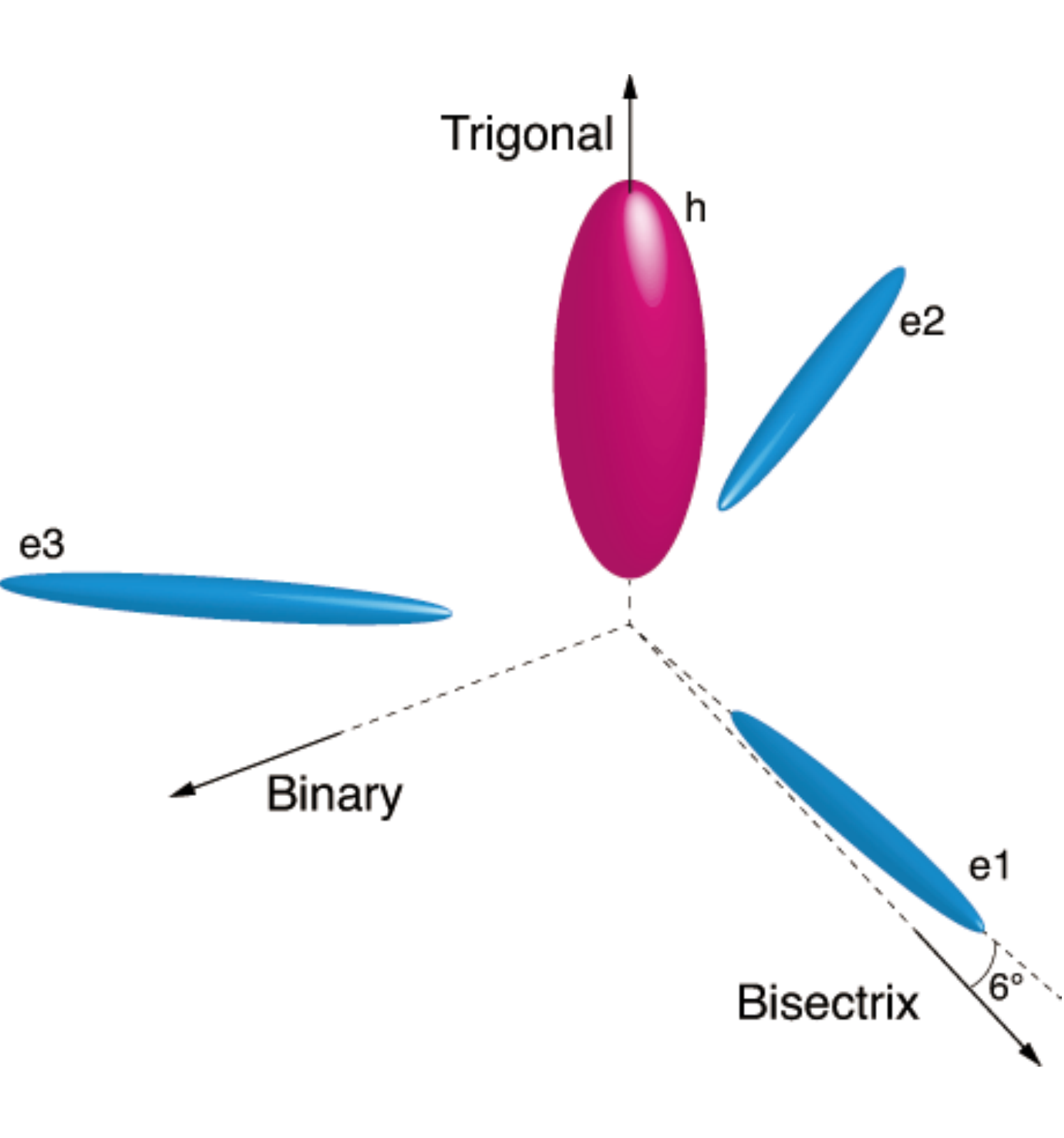}
		\caption{\label{Fig35} Illustration of Fermi surfaces of bismuth. One hole pocket locates at the $T$ point in the Brillouin zone along the trigonal axis. Three electron pockets locate at three equivalent the $L$ point with $2\pi/3$ interval. The electron pockets are almost parallel to the binary-bisectrix plane, but tilted slightly ($\sim 6^\circ$) in the trigonal direction.}
	\end{center}
\end{figure}

\begin{figure}
	\begin{center}
		\includegraphics[width=7cm]{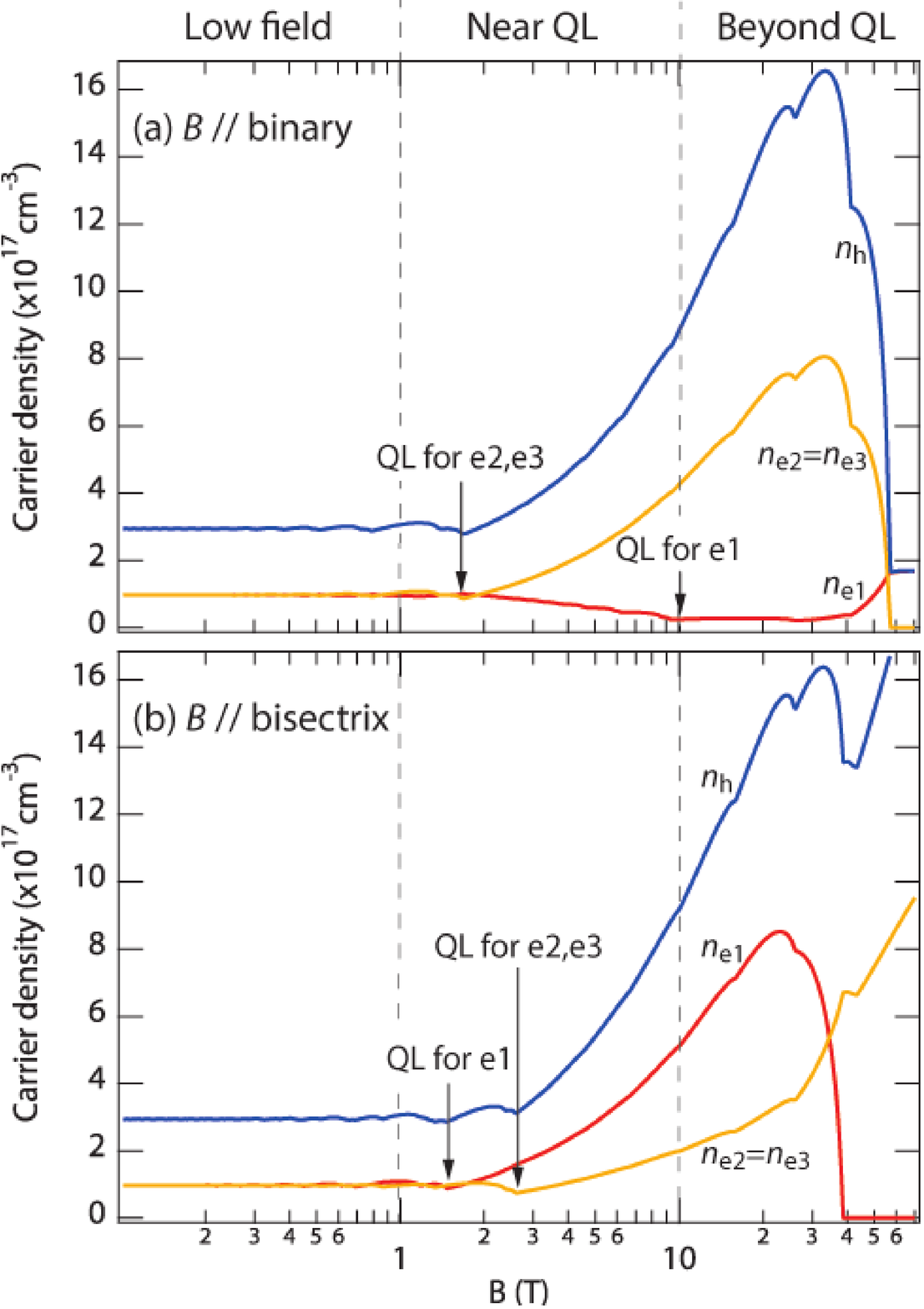}
		\caption{\label{Fig11} Carrier densities of electrons ($n_a$, $n_b$, $n_c$) and hole ($n_h$) in bismuth for (a) $B \parallel$ binary, and (b) $B \parallel$ bisectrix axis.
		There are three regions of magnetic field: (I) low filed limit, $B \lesssim 1$T; (II) near quantum limit (QL), $1 \lesssim B \lesssim 10$T; (III) beyond QL, $10\lesssim B$T.}
	\end{center}
\end{figure}
In bismuth, there are three electron valleys at the $L$ point and one hole valley at the $T$ point in the Brillouin zone (Fig. \ref{Fig35}). The three electron valleys are interchangeable upon a $2\pi/3$ rotation around the trigonal axis.
The valley degree of freedom in bismuth exhibit various intriguing aspects. They are classified into three regions according to the magnitude of magnetic field. They are (I) low field limit, (II) near the quantum limit (QL), and (III) beyond the QL. Here, QL state is the state where the whole carriers are confined into their lowest Landau level. The difference between the three regions are clearly seen in the field dependence of the carrier density shown in Fig. \ref{Fig11}.

\subsection*{(I) Low filed limit ($B\lesssim 1$ T)}
In the low field limit, the carrier density is almost constant with respect to the magnetic field, so that the distribution of carriers are equal among the electron valleys. However, the magnetoresistance exhibits a remarkable angular dependences, which is observable even at room temperature \cite{ZZhu2011}. The origin of such an outstanding angular oscillation is the large and anisotropic mobilities, and the fact that valleys, ellipsoidal pockets of the Fermi surface, do not lie parallel to each other. As a consequence, the contribution of each valley to the total conductivity strongly depends on the orientation of magnetic field. The angular dependence of magnetoresistance is complex, but can be explained almost perfectly on the basis of the semiclassical theory. Therefore, we can consider this field region as a semiclassical region. Nevertheless, one mysterious phenomena was found in this region: the breaking of valley-symmetry, where the three fold symmetry of the underlying crystal is lost.

\subsection*{(II) Near the quantum limit ($1$ T $\lesssim B \lesssim$ 10 T)}
Near the QL, the energy of carriers are clearly quantized into their Landau levels, so that we need to go beyond the semiclassical picture. 
The magnetic field at which the carrier attain the QL is different between valleys, since both cyclotron mass and g-factor are highly anisotropic in bismuth. Consequently, the difference between valleys becomes visible, resulting in the valley polarization, where the carriers distribute unevenly among three electron valleys. In this field region, the carrier distribution is not perfectly polarized but partially polarized, i.e., each valley has a finite carrier density (Fig. \ref{Fig11}), which should be distinguished from the complete valley-polarized state in the next field region.

\subsection*{ (III) Beyond the quantum limit ($B\gtrsim$ 10 T)}
The fate of electrons far beyond the QL is still unexplored, even though there are various proposals. Since the cyclotron mass of bismuth is very small $\sim 10^{-3}m_0$ ($m_0$ is the bare electron mass), the QL is easily achieved. For example, one of the electron valley attains the QL only with 1.5 T, and whole electron valleys attain it with 2.5 T for $\bB\parallel$ bisectrix (Fig. \ref{Fig11}). Therefore, bismuth is one of the best crystals to explore the physics far beyond the QL. So far, no phase transition and semimetal-semiconductor transitions has been detected up to 65 T for $\bB\perp $ trigonal and 90.5 T for $\bB \parallel$ binary axis. In stead of that, a  phenomena, which has never been encountered before in any other solids, was found; the 100\% valley-polarized state is attained by a strong magnetic field. This state is symbolically displayed as the abrupt drop in carrier density in Fig. \ref{Fig11}. Only by changing the direction of magnetic field, we can control the one- or two-valley emptying --- a new direction of controlling the valley degree of freedom.

In this review, we shall describe the properties of bismuth in three different regions of magnetic field. One would be able to learn a valuable lesson from an encyclopedic solid like bismuth.

\section{Magnetoresistance in multi-valley systems}\label{multivalley}
Here, we will briefly review the fundamental aspects of magnetoresistance based on the (semi-) classical picture. In overdoped cuprates the Fermi surface has a single component (a warped cylinder) and it has still a magnetoresistance. The same is true of many 2D organic system, even if its mobility is anisotropic. It is also well known that, when electrons and holes are compensated with each other, the magnetoresistance depends on the field as $\propto B^2$. On the other hand, it is less known that the magnetoresistance does not need the presence of carriers of both signs. When there are more than one anisotropic valley, the magnetoresistance is proportional to $B^2$ at low fields even if there is no compensated hole carrier. In the following, let us see these step by step.

Starting from the classical equation of motion in the external electric ($\bE$) and magnetic ($\bB$) field,
\begin{eqnarray}
	m^* \frac{d\bv}{d t}= -e \left(\bE + \bv \times \bB\right) -\frac{m^*}{\tau}\bv
\end{eqnarray}
($m^*$ is the effective mass, $e>0$ is the elementary charge and $\tau$ is the relaxation time),
the velocity of electrons at stationary states ($d\bv /dt=0$) is given as
\begin{equation}
	\bv = -\hmu \cdot \left( \bE + \bv \times \bB \right),
\end{equation}
by using electron mobility tensor $\hmu$.
Here, the anisotropy of both $\tau$ and $m^*$ are included in this mobility tensor.

\subsection{Single valley systems}
First, we consider the case only with a single valley.
When the magnetic field is along the $z$-direction, $\bB=(0, 0, B)$, the inplane components of velocity are
\begin{eqnarray}
		v_x =\frac{- \mu_x E_x + \mu_x \mu_y B E_y }
		{1+\mu_x \mu_y B^2},
		\\
		v_y =\frac{ -\mu_x \mu_y B E_x - \mu_y E_y}
		{1+\mu_x \mu_y B^2}.
\end{eqnarray}
The conductivity tensor, which is defined by
$	\bm{j}=-en  \bv = \hat{\sigma}\bE,
$
($n$ is a carrier density of electrons)
is obtained as
\begin{eqnarray}
	\hat{\sigma}(B) = \frac{n e}{1+ \mu_x \mu_y B^2}
	\left(
	\begin{array}{cc}
		\mu_x & - \mu_x \mu_y B
		\\
		+ \mu_x \mu_y B & \mu_y
	\end{array}
	\right).
	\label{MC1}
\end{eqnarray}
The magnetoresistance tensor $\hat{\rho}$ is then given by the inverse matrix of $\hat{\sigma}$ as
\begin{eqnarray}
	\hat{\rho}(B) = \frac{1}{n e \mu_x \mu_y}
	\left(
	\begin{array}{cc}
		\mu_y &  + \mu_x \mu_y B
		\\
		-\mu_x \mu_y B & \mu_x
	\end{array}
	\right).
	\label{magnetoresistance1}
\end{eqnarray}
It is clear from Eq. (\ref{magnetoresistance1}) that the transverse resistivity, $\rho_{xx}$ and $\rho_{yy}$, is independent from the magnetic field even if we take into account the anisotropy of mobility. This may be incompatible with one's intuition: the magnetoresistance is caused by the fact that the Lorentz force prevents electrons from running straight. This image is true for the transverse \emph{conductivity} of the system with a closed Fermi surface, where the transverse conductivity is reduced by the magnetic field. However, it is not so for the transverse \emph{resistivity}, where the field dependence is canceled out by the determinant of the conductivity tensor. Like this, in many cases, it is not straightforward to understand the magnetoresistance based on the intuitive picture.

The Hall resistivity is obtained as
\begin{eqnarray}
	\rho_{yx} (B)=-\frac{B}{ne},
\end{eqnarray}
and the Hall coefficient, $R_{\rm H}\equiv \rho_{yx}(B)/B$, becomes
\begin{eqnarray}
	R_{\rm H}(B)=-\frac{1}{ne},
\end{eqnarray}
which enables us to evaluate the carrier density and the sign of the carrier only by measuring $R_{\rm H}$.

The results shown above are rigorous within a classical (and even semiclassical at zero temperature) picture, and they are applicable for any amplitude of magnetic field.

\subsection{Semimetallic systems}
Next, we consider the case where both electron and hole carriers coexist, i.e., the semimetallic systems. We express the mobility tensor of hole carries by $\hnu$. The conductivity tensor for holes can be obtained just by replacing $B \to -B$ in Eq. (\ref{MC1}). The total conductivity tensor is given by a summation of the conductivity of each carriers as
\begin{eqnarray}
	\hat{\sigma}_{\rm e+h}(B)&=\hat{\sigma}_{\rm e} (B) +\hat{\sigma}_{\rm h}(B)
	\nonumber\\
	&=\left(
	\begin{array}{cc}
	\sigma_{\rm e1} + \sigma_{\rm h1} & -\sigma_{\rm e2}+\sigma_{\rm h2} \\
	\sigma_{\rm e2} - \sigma_{\rm h2} & \sigma_{\rm e1} + \sigma_{\rm h1}
	\end{array}
	\right),
\end{eqnarray}
where
\begin{eqnarray}
	\sigma_{\rm e1} &= \frac{ne\mu_x}{1+\mu_x \mu_y B^2},
	\quad
	\sigma_{\rm e2} &= \frac{ne\mu_x \mu_y B}{1+\mu_x \mu_y B^2},
	\\
	\sigma_{\rm h1} &= \frac{pe\nu_x}{1+\nu_x \nu_y B^2},
	\quad
	\sigma_{\rm h2} &= \frac{pe\nu_x \nu_y B}{1+\nu_x \nu_y B^2}.
\end{eqnarray}
($p$ is a carrier density of holes.)
The transverse magnetoresistance is obtained as
\begin{eqnarray}
	\rho_{xx}^{\rm e+h}=\frac{\sigma_{\rm e1} + \sigma_{\rm h1}}
	{\left(\sigma_{\rm e1}+\sigma_{\rm h1}\right)^2 +\left(\sigma_{\rm e2}-\sigma_{\rm h2}\right)^2}.
\end{eqnarray}
At high fields of $\mu_x \mu_y B^2 \gg 1$ and $\nu_x \nu_y B^2 \gg 1$, the magnetoresistivity becomes
\begin{eqnarray}
	\rho_{xx}^{\rm e+h}(B)\simeq \frac{1}{e}
	\frac{\displaystyle \left( \frac{n}{\mu_y}+\frac{p}{\nu_y}\right)B^2}
	{\displaystyle \left( \frac{n}{\mu_y}+\frac{p}{\nu_y}\right)^2 + (n-p)^2 B^2}.
\end{eqnarray}
For $n\neq p$, the magnetoresistance saturates at around a characteristic field:
\begin{eqnarray}
B^*=\frac{n\nu_y +p\mu_y}{|n-p|\mu_y \nu_y}.
\end{eqnarray}
For $n=p$, however, the magnetoresistance never saturates and exhibits $B^2$ dependence as
\begin{eqnarray}
	\rho_{xx}^{\rm e+h}(B)\simeq \frac{B^2}{ne}\frac{\mu_y \nu_y}{\mu_y + \nu_y}.
\end{eqnarray}
The Hall resistivity at high fields is also obtained in the form
\begin{eqnarray}
	\rho_{yx}^{\rm e+h}(B)\simeq
	-\frac{B}{e}\frac{(n-p)B^2}
	{\displaystyle \left( \frac{n}{\mu_y}+\frac{p}{\nu_y}\right)^2
	+(n-p)^2B^2}.
\end{eqnarray}
For $B \gg B^*$, the Hall coefficient, $R_{\rm H}\equiv \rho_{yx}(B)/B$, becomes
\begin{eqnarray}
	R_{\rm H}(B\gg B^*)=-\frac{1}{(n-p)e}.
\end{eqnarray}
It should be noted here that the above form of $R_{\rm H}$ is valid only in the ``high field" limit of $B\gg B^*$, which is in contrast to the single carrier systems where $R_{\rm H}=1/ne$ holds for any magnitude of the field.

\subsection{Multi-valley systems}
\begin{figure}
	\begin{center}
		\includegraphics[width=8cm]{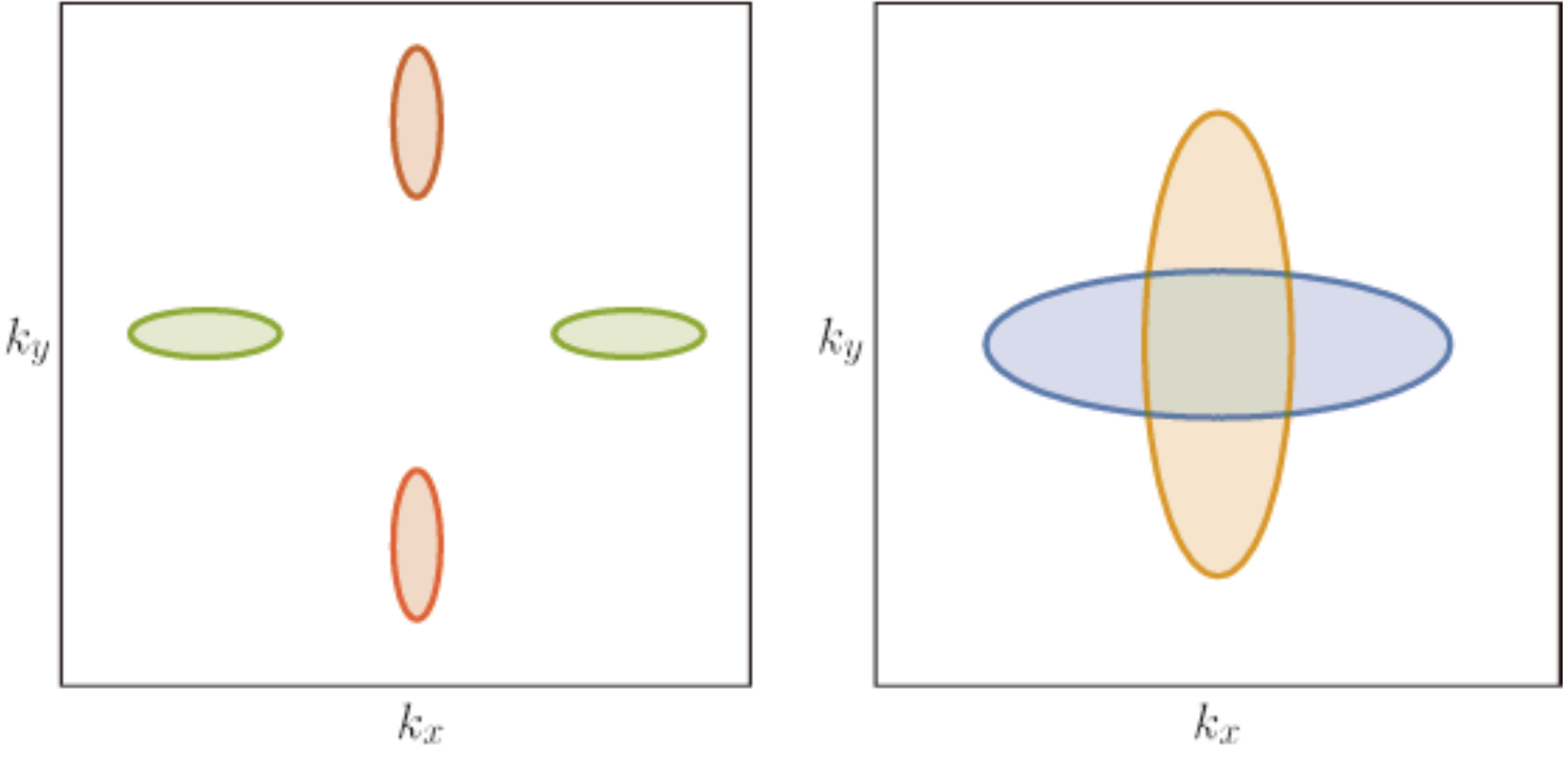}
		\caption{\label{Fig21} Examples of the Fermi surfaces of two kinds of valley, where the mobility of one valley is obtained by 90$^\circ$ rotation of the other. (Left) Four valleys and (Right) two valleys system. In these cases, it is easily shown that the magnetoresistivity is proportional to $B^2$ in the low field.
		}
	\end{center}
\end{figure}
It is less known that the multi-valley systems (only one type of carrier, but with more than one valley) also exhibit the magnetoresistance. Here, as a simple example, we consider the system with two anisotropic electron valleys, assuming that the mobility of one valley is obtained by 90$^\circ$ rotation of the other valley (Fig. \ref{Fig21}).
The total magnetoresistance tensor is obtained in the form
\begin{eqnarray}
	\hat{\rho}_{1+2} (B)
	&=\frac{1}{ne}
	\frac{1+\mu_x \mu_y B^2}{(\mu_x + \mu_y)^2+4\mu_x^2  \mu_y^2 B^2}
	\nonumber\\
	&\times
	\left(
	\begin{array}{cc}
		\mu_x + \mu_y & 2 \mu_x \mu_y B
		\\
		-2 \mu_x \mu_y B & \mu_x + \mu_y
	\end{array}
	\right),
\end{eqnarray}
which clearly indicates that $\rho_{xx}$ depends on the magnetic field. In the weak field limit,
\begin{eqnarray}
	\frac{\rho_{xx}(B)-\rho_{xx}(0)}{\rho_{xx}(0)}\simeq
	\frac{ (\mu_x-\mu_y)^2}{(\mu_x + \mu_y)^2}\mu_x \mu_yB^2 .
\end{eqnarray}
Therefore, the multi-valley systems exhibit $B^2$ dependence in $\rho_{xx}$, and it is proportional to the anisotropy and the magnitude of mobilities.

\section{Tuning contribution of valleys I: Low field limit}\label{Low field limit}
\subsection{Angular dependence of magnetoresistance in bismuth}

What distinguishes bismuth from any other solids is the fact that the angular oscillations of the magnetoresistance are visible even at room temperature for a field as low as 0.7 T \cite{ZZhu2011b}. 
This is a consequence of the large mobility of electrons ($\sim 10^4$ cm$^2$V$^{-1}$s$^{-1}$= 1 T$^{-1}$ at room temperature) and highly anisotropic mass (0.26 $m_{\rm e}$ along bisectrix and 0.0011 $m_{\rm e}$ along binary). No solid other than bismuth is currently known to present such properties.
\begin{figure*}
	\begin{center}
		\includegraphics[width=14cm]{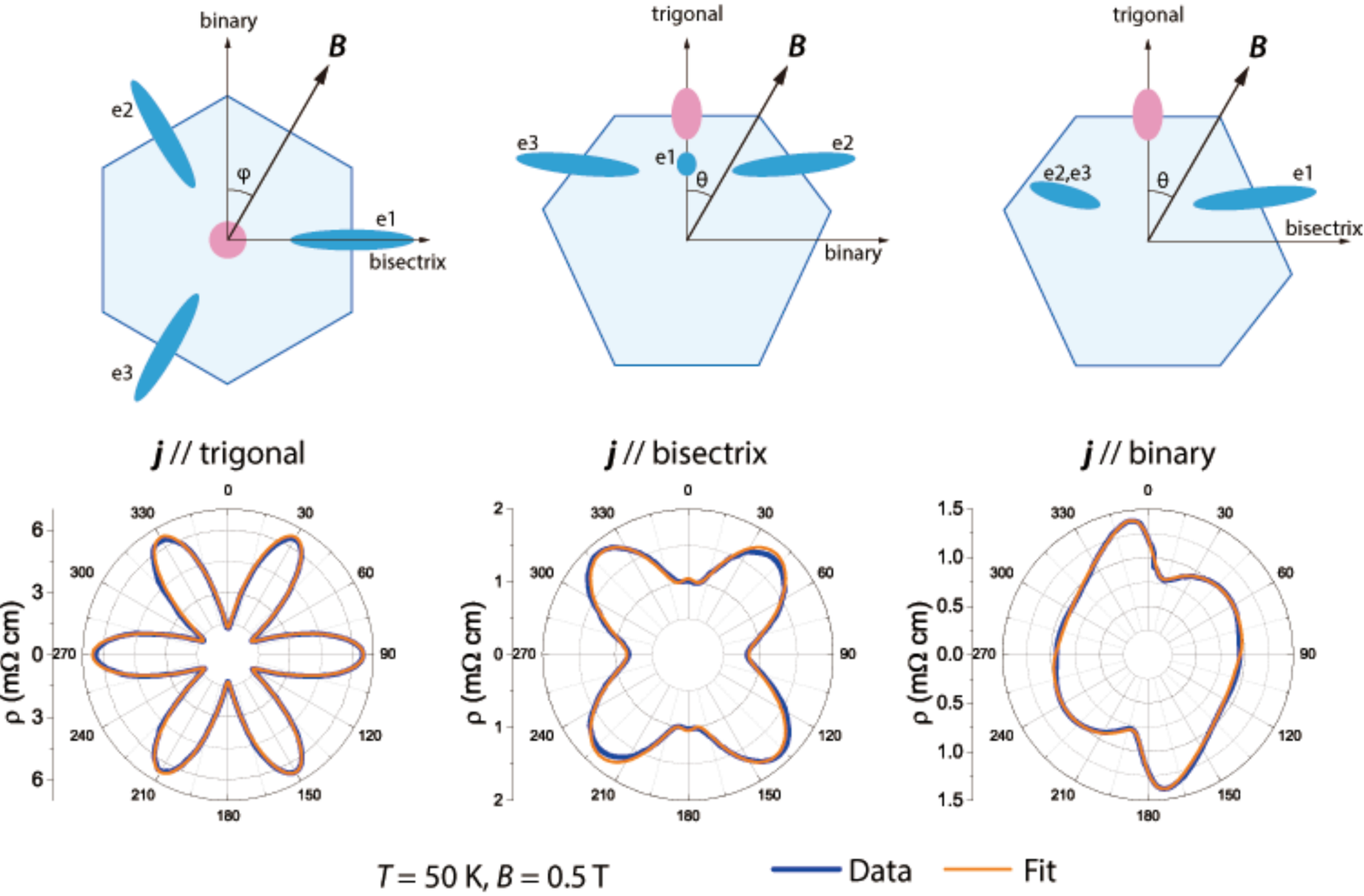}
		\caption{\label{Fig31} Angle dependence of transverse magnetoresistance in polar plots for three perpendicular planes at a typical temperature 30 K and field 0.5 T. The electric current is applied along the axis perpendicular to the rotation plane of the magnetic field. The field rotates in (a) binary-bisectrix, (b) trigonal-binary, and (c) trigonal-bisectrix plane. Upper panels show the projection of the Brillouin zone, the hole and electron pockets of the Fermi surface in each configuration.}
	\end{center}
\end{figure*}

Because of these specific properties, bismuth is one of the most suitable solids to investigate the angular dependence of magnetoresistance. Figure \ref{Fig31} shows polar plots of transverse magnetoresistance ($T=30$ K, $B=0.5$ T) for three different direction of electric current: (a) $\bj \parallel $ trigonal-, (b) $\bj \parallel $ bisectrix-, (c) $\bj \parallel $ binary-axis. The magnetic field is rotated in the plane perpendicular to the applied current.
When $\bj \parallel $ trigonal axis, the magnetoresistance exhibits a sixfold angular dependence, reflecting the $C_{\rm 3v}$ symmetry of the crystal [see the upper panel of Fig. \ref{Fig31} (a)]. It takes its maximum (minimum) values when $\bB \parallel$ bisectrix (binary) axis. In this plane, the contribution from hole valley is isotropic, so that the angular dependence originates from the anisotropy of three electron valleys.
When $\bj \parallel $ bisectrix axis, it is mirror symmetric as $\rho_{22}(\theta)=\rho_{22}(-\theta)$ and inversion symmetric as $\rho_{22}(\theta)=\rho_{22}(\theta + \pi)$. [cf. the upper panel of Fig. \ref{Fig31} (b).]
When $\bj \parallel$ binary axis, only the inversion symmetry is kept. For $\bj \parallel$ bisectrix- and binary-axes, the hole contribution becomes the largest when $\bB\parallel$ trigonal axis. The total magnetoconductivity is determined by the summation of the contributions from three electron and one hole pockets. As a result, the total magnetoresistance peaks at intermediate angles off the high-symmetry axes. Although it is too complex to evaluate the peak and dip positions only from the simple estimation, they can be explained perfectly if we compute $\rho(\theta)$ based on the one-particle semiclassical theory in the following.

\subsection{Semiclassical theory of angular dependence of magnetoresistance}

In Sec. \ref{multivalley}, we saw that the magnetoresistance depends on the magnetic field when the system is semimetallic, OR, has more than one anisotropic valleys. In the case of bismuth, it is semimetallic AND has three anisotropic electron valleys, each of which gives different contribution to the magnetoresistance. They are very complex, but understandable. Here, we first introduce more complete semiclassical theory of magnetoresistance for the multivalley and semimetallic systems. Then, we see how experiments can be interpreted almost perfectly by the one-particle picture based on the semiclassical theory.

For the calculation of the angular dependence of magnetic field, it is very convenient to express the magnetic field $\bB=(B_x, B_y, B_z)$ in the tensor form:
\begin{eqnarray}
	\hB = \left(
	\begin{array}{ccc}
		0 & -B_z & B_y\\
		B_z & 0 & -B_x \\
		-B_y & B_x & 0
	\end{array}
	\right).
\end{eqnarray}
By using this tensor and the mobility tensor, the current can be expressed as
\begin{eqnarray}
	\bj = ne \hmu \cdot \bE + \hmu\cdot (\hB \cdot \bj ).
\end{eqnarray}
Then, we obtain the conductivity tensor for arbitrary orientation and magnitude of magnetic field in the form:
\begin{eqnarray}
	\hat{\sigma}(\hB) &= &ne \left[ \hmu^{-1}- \hB\right]^{-1}.
	\label{conductivity}
\end{eqnarray}
(Again, we can obtain the conductivity for holes just by replacing $\hB \to -\hB$.)
One can obtain the same results based on the Boltzmann equation in the zero temperature limit \cite{Abeles1956,Mackey1969}. The formula (\ref{conductivity}) is extremely general, so that we can apply it to various systems even under a high magnetic field within the semiclassical theory. (It was shown that the the result obtained by Eq. (\ref{conductivity}) perfectly agrees with those by Kubo formula except for the quantum oscillations \cite{Owada2017}.)

By the careful measurements on the fermiology, it is well established that the mobility tensor of one electron valley (valley e1) has the form
\begin{eqnarray}
	\hmu^{\rm e1} = \left(
	\begin{array}{ccc}
	\mu_1 & 0 & 0 \\
	0 & \mu_2 & \mu_4 \\
	0 & \mu_4 & \mu_3
	\end{array}
	\right),
	\label{mua}
\end{eqnarray}
where the coordinates 1, 2, and 3 correspond to  binary, bisectrix, and trigonal axes, respectively.
By substituting Eq. (\ref{mua}) into Eq. (\ref{conductivity}), the conductivity tensor of valley e1 is obtained as \cite{Aubrey1971}
\begin{eqnarray}
	\sigma_{11}^{\rm e1} &= \left( \mu_1 + d B_1^2 \right)g^{\rm e1},
	\label{s11}
	\\
	\sigma_{22}^{\rm e1} &= \left( \mu_2 + d B_2^2 \right)g^{\rm e1},
	\\
	\sigma_{33}^{\rm e1} &= \left( \mu_3 + d B_3^2 \right)g^{\rm e1},
	\label{s33}
	\\
	\sigma_{12}^{\rm e1} & = \left\{ \mu_1 \mu_4 B_2 - \mu_1 \mu_2 B_3  +d B_1 B_2 \right\}g^{\rm e1},
	\\
	\sigma_{23}^{\rm e1} & = \left\{ \mu_4 -\mu_2 \mu_3 +\mu_4 ^2 B_1 + d B_2 B_3  \right\}g^{\rm e1},
	\\
	\sigma_{31}^{\rm e1} & = \left\{ -\mu_1 \mu_3 B_2 +\mu_1 \mu_4 B_3 + dB_3 B_1 \right\}g^{\rm e1},
\end{eqnarray}
where
\begin{eqnarray}
	g^{\rm e1} &= ne\bigl\{1+(\mu_2 \mu_3 -\mu_4^2)B_1^2 +\mu_1\mu_3 B_2^2 + \mu_1 \mu_2 B_3^2
	\nonumber\\&
	-2\mu_1 \mu_4 B_2 B_3 \bigr\}^{-1},
\end{eqnarray}
	and
\begin{eqnarray}
		d=\mu_1 \mu_2 \mu_3 -\mu_1 \mu_4^2
		\label{det}
\end{eqnarray}
is the determinant of $\hmu^{\rm e1}$.
The conductivity tensor of valley e2 and e3 are also obtained in the same manner. Their mobilities  are given by $\hmu^{\rm e2, e3}=R_3^{-1}(\pm 2\pi/3)\cdot \hmu^{\rm e1} \cdot R_3(\pm 2\pi/3)$, where $R_3 (\theta)$ is the rotation matrix by an angle $\theta$ about the trigonal axis.
The mobility tensor of hole at the $T$ point has the form
\begin{eqnarray}
	\hat{\nu}=\left(
	\begin{array}{ccc}
	\nu_1 & 0 & 0 \\
	0 & \nu_1 & 0 \\
	0 & 0 & \nu_3
	\end{array}
	\right).
\end{eqnarray}
Then the conductivity tensor for hole, $\sigma_{\mu \nu}^{h}$, can be obtained by letting $\mu_1, \mu_2 \to \nu_1$, $\mu_3 \to \nu_3$, and $\mu_4 \to 0$ with $B_i \to -B_i$ in Eqs. (\ref{s11})-(\ref{det}).
The total conductivity is given by
$
	\sigma_{\mu\nu}= \sigma_{\mu \nu}^h + \sum_i \sigma_{\mu \nu}^{{\rm e}i},
$
and the resistivity tensor is
$
	\hat{\rho}=\hat{\sigma}^{-1}.
$

By using these formulae of magneto-conductivity, the angular dependence of magnetoresistance in bismuth is perfectly explained as is shown in Fig. \ref{Fig31} \cite{ZZhu2011b,Collaudin2015}. Here, the mobilities $\mu_{1,2,3}$ and $\nu_{1,3}$ are determined so as to fit the experimental results, while the carrier density and the tilt angle $\phi$ are fixed at $n_{\rm e}=n_{\rm h}=3\times 10^{17}$ cm$^{-3}$ \cite{Bhargava1967} and $\phi=6.8^\circ$ \cite{Bhargava1967,ZZhu2011}, respectively. (The tilt angle is related to the mobility tensor by $\phi=\arctan[2\mu_4/(\mu_2 -\mu_3)]/2$.)
It should be stressed here that a unique set of mobilities $\mu_{1,2,3}$ and $\nu_{1,3}$ can give qualitatively good agreement with experiments for whole perpendicular planes, indicating the validity of the current semiclassical theory of magnetoresistance.

\subsection{Temperature dependence of mobility tensors}

\begin{figure}
	\begin{center}
		\includegraphics[width=6cm]{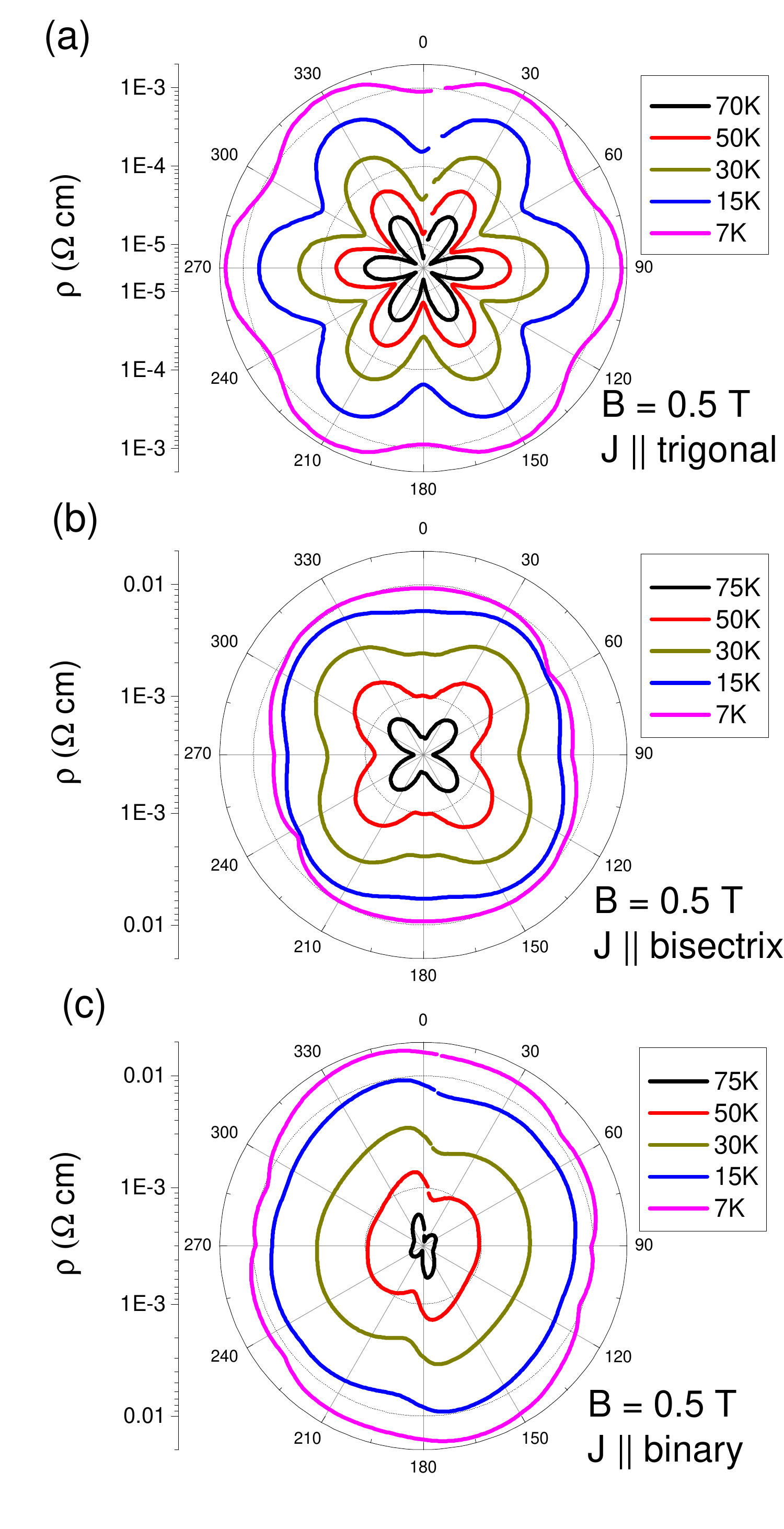}
		\caption{\label{Fig33} Temperature evolutions of the angular dependence of magnetoresistance at $B=0.5$ T for three perpendicular rotation planes. The shape of the angular dependence changes qualitatively with decreasing temperature, but they are interpreted based on the one-particle picture of semiclassical theory. The qualitative change is just because the different temperature dependence between the mobilities.
		}
	\end{center}
\end{figure}
The evolutions of the angle dependence of the transverse magnetoresistance with decreasing temperature are shown in Fig. \ref{Fig33}.
The shape of polar plot seems to be changed ``qualitatively" by decreasing temperature. For example, when $\bj \parallel$ bisectrix, it looks like a four-leaf clover at high temperatures, while it is almost circular at low temperatures. However, this does not mean neither the qualitative transition in the electronic state nor the broken of one-particle semiclassical picture. At each temperature, the one-particle semiclassical theory can fit the experimental data as is shown in Fig. 5 of Ref. \cite{Collaudin2015}. The apparent qualitative change is just because of the fact that the relative contributions between the mobility components is changed, since each mobility shows different temperature dependence, especially in their slope of $T^2$, as shown in the following.

\begin{figure}
	\begin{center}
		\includegraphics[width=7cm]{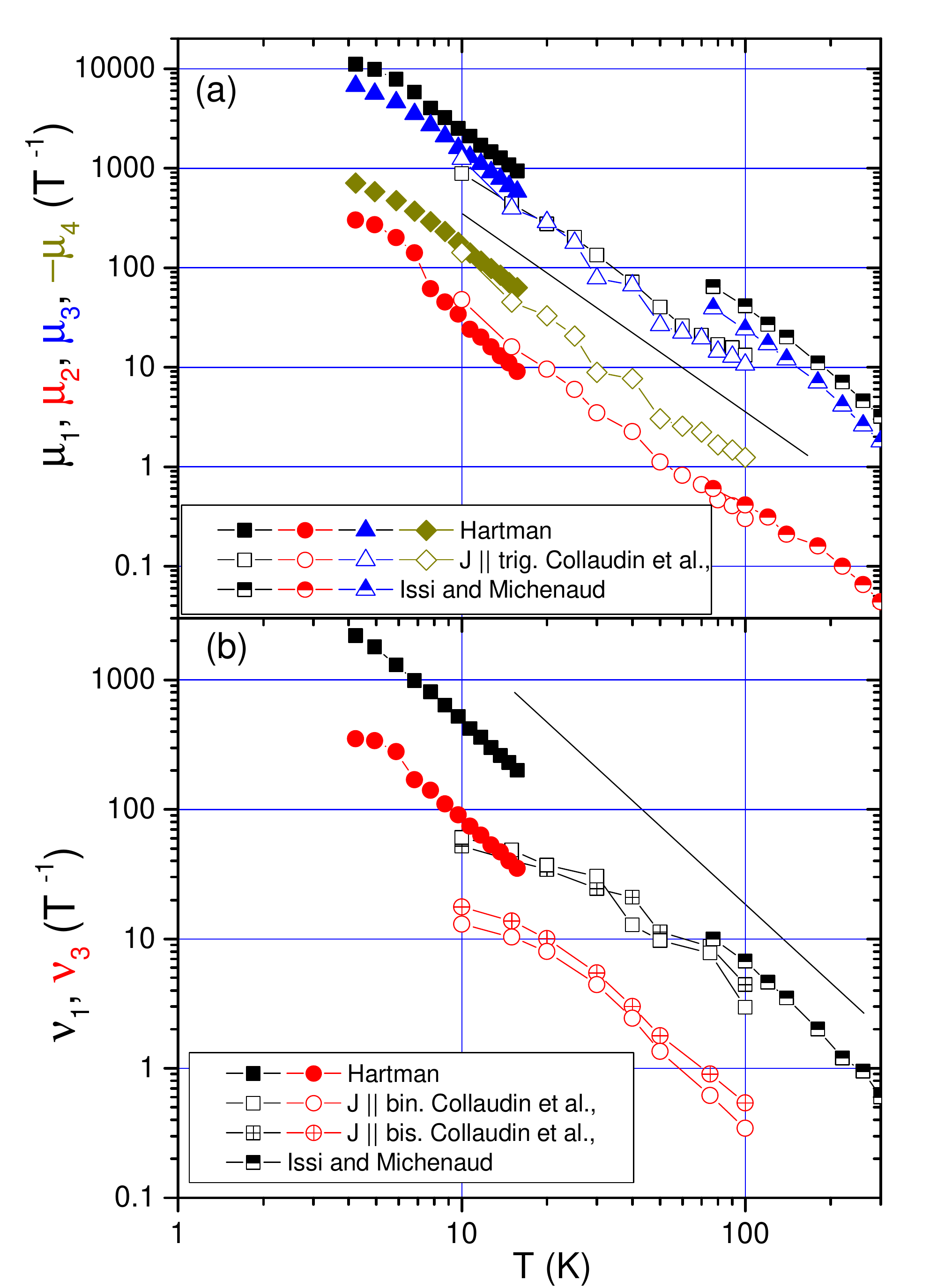}
		\caption{\label{Fig34} Temperature dependence of the components of (a) the electron mobility tensor $\mu_i$ for a current along trigonal, and (b) the hole mobility tensor $\nu_i$ by fitting the angle-dependent magnetoresistance data for a current along binary or bisectrix axis at a field of 0.5 T. For comparison, the results reported by Hartman below 15 K \cite{Hartman1969} and by Michenaud and Issi above 77 K \cite{Michenaud1972} are shown.}
	\end{center}
\end{figure}

Figure \ref{Fig34} shows the temperature dependence of the mobility components for (a) electrons and (b) holes extracted by fitting the angular dependence of magnetoresistance at $B=0.5$ T. The zero-field limit values of mobility components obtained by Hartman ($T<15$ K) \cite{Hartman1969} and by Michenaud and Issi ($T>77$ K) \cite{Michenaud1972} from the galvanometric coefficients are also shown in Fig. \ref{Fig34}.
One can see from Fig. \ref{Fig34} that the mobility of bismuth is extremely high: $\mu_1$ and $\mu_3$ is as large as 10$^3$ T$^{-1}$ (10$^7$ cm$^2$V$^{-1}$s$^{-1}$) at 10 K, and 10$^4$ T$^{-1}$ (10$^8$ cm$^2$V$^{-1}$s$^{-1}$) at 4 K. This is larger than the super-clean two-dimensional electron gas ($\sim 3.5\times 10^7$ cm$^2$V$^{-1}$s$^{-1}$ at $T\lesssim 1$ K \cite{Umansky2009}) and much larger than the carbon nanotube ($\sim 10^5$ cm$^2$V$^{-1}$s$^{-1}$ at room temperature \cite{Durkop2004}) or the graphene ($\sim 2\times10^5$ cm$^2$V$^{-1}$s$^{-1}$ at $T\sim 5$ K \cite{Bolotin2008}). It should be emphasized here that electrons in bismuth are far more mobile than carriers in any other three-dimensional solid \cite{TLiang2014}.

As was pointed out by Hartman \cite{Hartman1969}, each mobility exhibit $\mu_i \propto T^{-2}$ for a wide rage of temperature below room temperature, indicating that the electron-electron scattering is the dominant process for the relaxation time \cite{Baber1937,Abrikosov1959}. On the other hand, at low enough temperatures $T\lesssim 5$ K, each mobility becomes almost independent from temperature, suggesting the dominant scattering process is changed to the impurity scattering with decreasing temperature.

It is interesting to note the consistency and the difference between various studies. The mobility of holes in the samples studied by Collaudin et al. \cite{Collaudin2015} saturates at a temperature well above those studied by Hartmann et al \cite{Hartman1969}. The RRR of the sample studied in the latter case was much higher. Therefore, the comparison suggests that increases in disorder affects the mobility of hole-like carriers more than electron-like carriers.

\section{Tuning contribution of valleys II: Near the quantum limit}

\subsection{Angle resolved Landau spectrum}
Bismuth is known as the solid in which the quantum oscillation was discovered \cite{Shubnikov1930,Haas1930}. It is mainly because of the extremely light cyclotron mass ($\sim 10^{-3} m_0$) and long mean free path ($\sim 0.3$ mm) of bismuth \cite{Fuseya2015}.
In spite of such an easy detection of the quantum oscillations, the angular dependence of the quantum oscillation in bismuth is very complex and it is not so straightforward to analyze. Three electrons and one hole pockets give different contribution to the spectrum. The Fermi energy drastically changes with magnetic field near the QL. This is due to the restriction of the charge neutrality, where the carrier numbers of electron and holes are kept to be equal, and the large difference of the cyclotron mass between electrons and holes (e.g., $m_{\rm c}=0.00189$, $M_{\rm c}=0.221$ for $\bB\parallel$ bisectrix, cf. Table \ref{Table2}).

What makes the angle-resolved Landau spectrum of bismuth more intriguing is the angular dependence of the Zeeman splitting. The Zeeman energy of bare electrons in vacuum is isotropic with respect to the orientation of the magnetic field, and its magnitude, the g-factor is $g=2$. These are the automatic consequences of Dirac theory in vacuum. However, under a crystalline potential and with a sizable spin-orbit interaction, the g-factor is enhanced and becomes anisotropic. Therefore, from the analysis of the angle-resolved Landau spectrum, we can obtain the information of not only the Fermi surface, but also the spin-orbit coupling in crystal.

\begin{figure}
	\begin{center}
		\includegraphics[width=7cm]{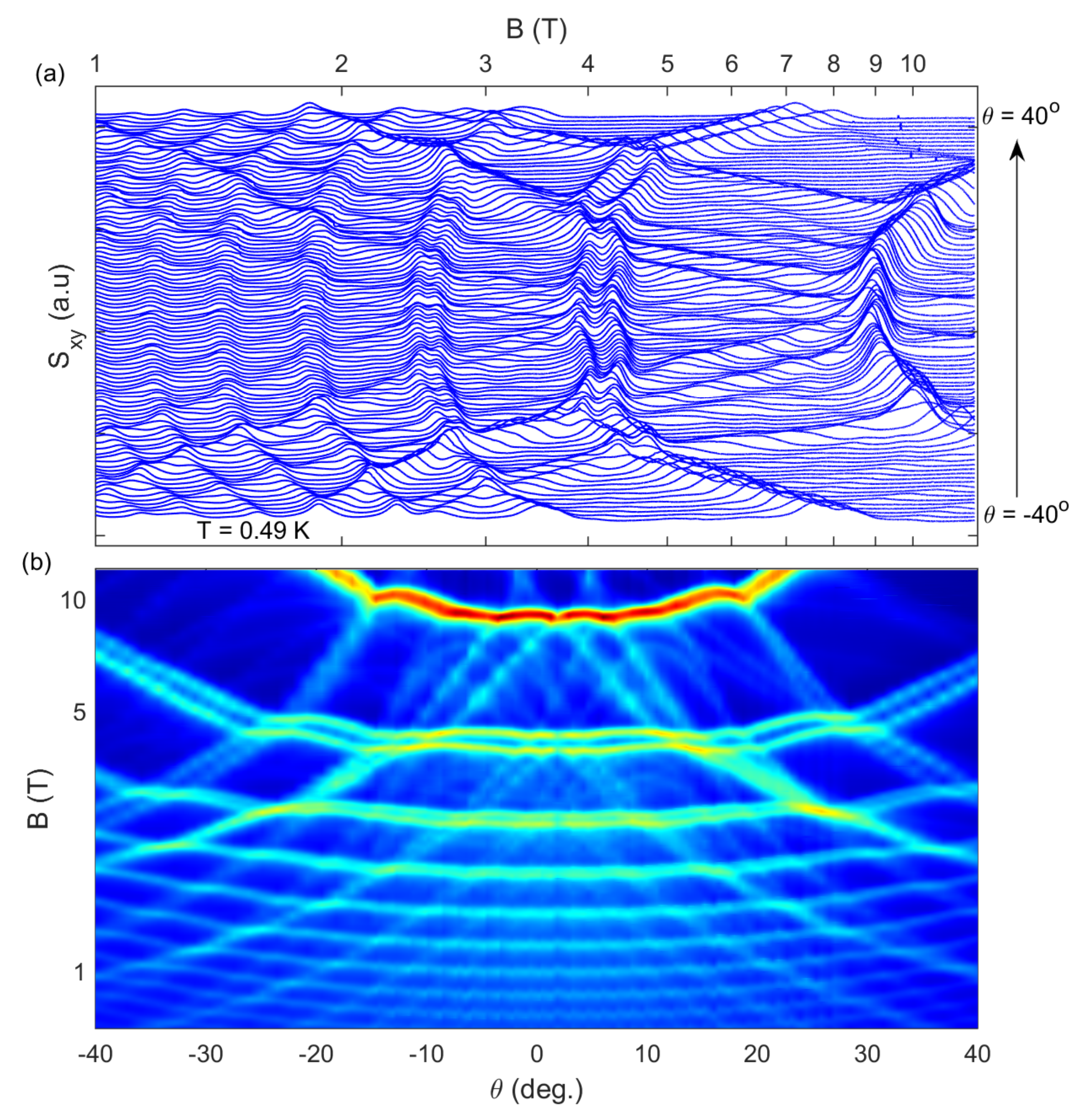}
		\caption{\label{Fig46} (a) Field dependence of the Nernst signal $S_{xy}$ with the magnetic field as the field rotates in the trigonal-binary plane from $-40^\circ$ to $40^\circ$ at $T=0.49$ K. Curves are shifted for clarity. (b) Color map of the same data. The bright lines track the angular evolution of the Nernst peaks where the Landau level intersects the chemical potential. The quasi-horizontal lines originate from the contribution of holes.}
	\end{center}
\end{figure}
Figure \ref{Fig46} shows a typical data set of the angle-dependence of Nernst response. The upper panel shows shifted Nernst signal as a function of magnetic field rotating along trigonal-binary plane. 
The lower panel shows the color map of the same data. The bright lines correspond to the position where the Nernst response is peaked, and they clearly display the Landau spectrum. The angle dependence make it easier to identify the spectrum, while it is hard for the quantum oscillation data with one orientation to distinguish the electron and hole spectrum. The quasi-horizontal lines can be attributed to the spectrum due to the hole pocket, since the longer axis of the hole ellipsoid is along the trigonal axis, and the cross section of this ellipsoid increases as the field is tilted. On the other hand, the quasi-perpendicular lines correspond to the three electron pockets, of which longer axis is almost perpendicular to that of holes.

\subsection{Effective model}\label{Effective model}

For the detailed analysis of the angle resolved Landau spectrum, we need to introduce an effective model for the energy under the magnetic field. Because of the narrow gap character of electrons, a model only with a single band is not enough to give a satisfactory agreements with experiments. To take into account the system with coupled bands, $\kp$ theory is quite powerful. Cohen and Blount applied $\kp$ theory to the conduction and valence bands at the $L$ point in bismuth considering the spin-orbit coupling in a fully relativistic way \cite{Cohen1960}. They showed that the effective g-factor is exactly given by $g^*=2m_0/m_c$, where $m_c$ and $m_0$ being the cyclotron effective mass and the bare electron mass, respectively. The energy is then obtained as
$
E_{n, \sigma}^{\rm e} =\pm\sqrt{ \D^2 + 2\D\left\{ \left(n+1/2 +\sigma/2\right )\hbar \omega_c + \hbar^2 k_z^2/2m_z \right\} },
$
where $\sigma=\pm 1$.
This is exactly the same situation of Dirac electrons as was shown by Wolff \cite{Wolff1964}. Therefore, according to this two-band analysis, the Landau levels should be doubly degenerated and there should be no spin splitting.

\begin{figure}[htb]
	\begin{center}
		\includegraphics[width=7cm]{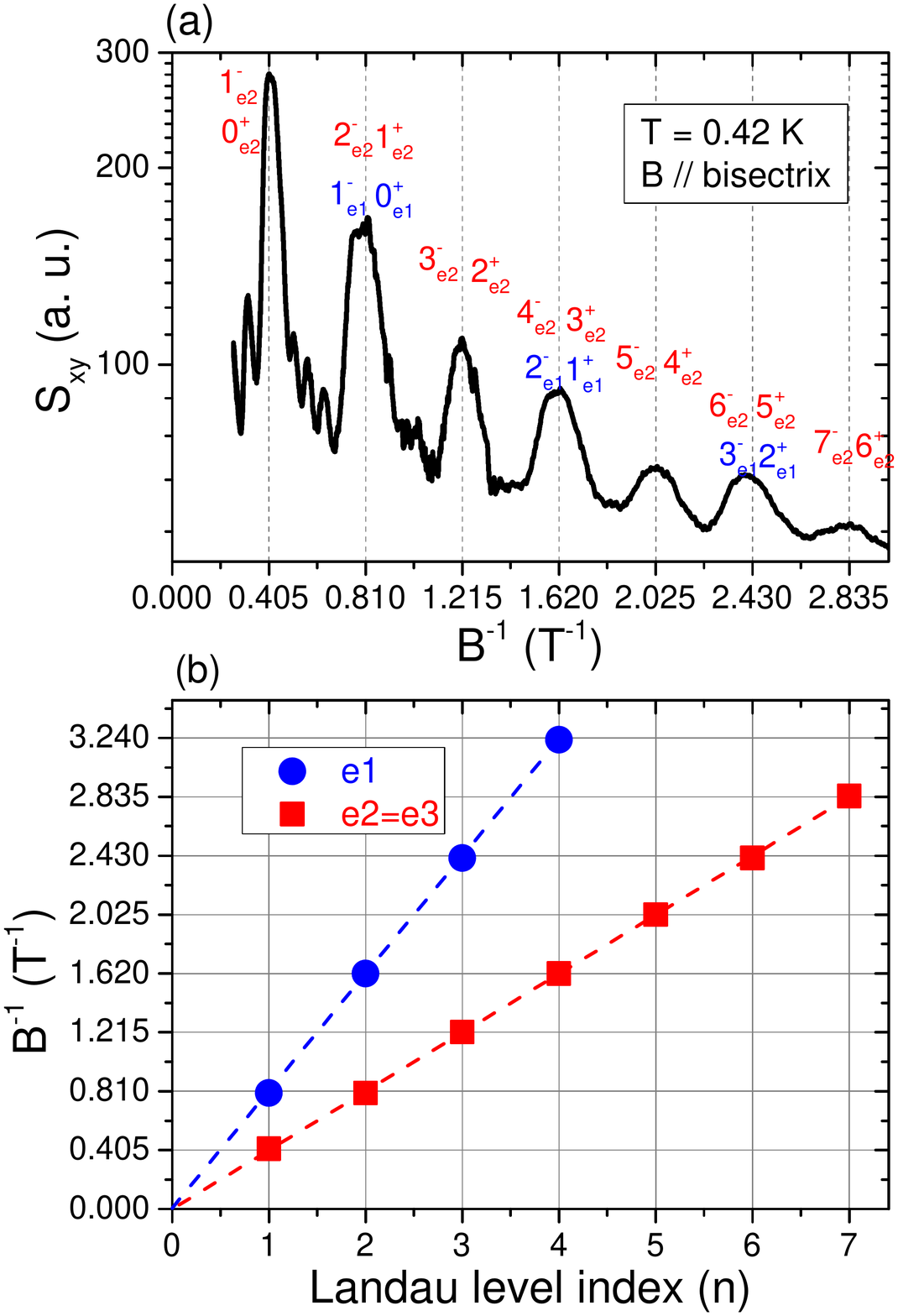}
		\caption{\label{Fig44} (a) Nernst voltage as a function of $B^{-1}$ with the field along the bisectrix axis. The quantum oscillation period of the electron valleys e2, e3 is 0.405 T$^{-1}$ which is twice of the electron valley e1 (cf. Fig \ref{Fig35}). When $B^{-1}$ is an even (odd) multiple of 0.405 T$^{-1}$, a Nernst peak is sixfold (fourfold) degenerate. Thus, the amplitude of the Nernst peaks with the even multiple is higher than that with odd multiple. (b) Value of $B^{-1}$ of the Nernst peaks as a function of their Landau index for the three valleys. Dirac spectrum leads to their vanishing intercepts.}
	\end{center}
\end{figure}
Figure \ref{Fig44} (a) is the plot of the Nernst response for $ \bB \parallel$ bisectrix ($0\le B^{-1} \le 3$ T$^{-1}$). It is clear that the peak is periodic with a periodicity of $0.405 \pm 0.005$ T$^{-1}$. In this direction, the spin splitting for electrons is zero, which is proved by the two crossing spectrum in the field rotation around the bisectrix axis. (This should be distinguished from the experiments where the spin splitting is not observed just because of its less resolution.) 
The magnetic fields at which the Nernst peaks locate are plotted as a function of their Landau level index in Fig. \ref{Fig44} (b). The ``zero" intercept of the plot of Fig. \ref{Fig44} (b) indicates that the energy of the electrons is given by that of Dirac electrons as $E_F^2/B=(n+1/2\pm1/2)\hbar e/m_c$.

For the other orientation of the magnetic field, however, finite spin splittings have been obseved by experiments \cite{Kunzler1962,Smith1964}. In order to recover the spin splitting, Lax {\it et al.} introduced a phenomenological g-factor in the two-band $\kp$ model as
$
E_{n, \sigma}^{\rm e} =\pm\sqrt{ \D^2 + 2\D\left\{ \left(n+1/2\right )\hbar \omega_c + \hbar^2 k_z^2/2m_z + \sigma g^*\mu_{\rm B} B/2\right\} },
$
where $\mu_{\rm B}=e\hbar/2m_0$ is the Bohr magneton. Smith, Baraff and Rowell used this model to analyze their angle-resolved Shubnikov-de Haas oscillations, and obtained a good account of the experiment, especially when the field is oriented close to the trigonal axis \cite{Smith1964,Sharlai2009}.

However, there is a serious problem in the two-band model of Lax. If the contribution from the term $-g^*\mu_{\rm B}B/2$ is negatively large enough, the energy becomes imaginary, which is, of course, physically incorrect. That's the reason why it is applicable only the case with the low field oriented close to the trigonal axis, where $g^*$ is small \cite{Smith1964,Sharlai2009}. This problem was removed by Baraff, who obtained microscopically the $g^*$-term taking into account the bands other than the two-band model by means of perturbation theory \cite{Baraff1965}. Unfortunately, the Baraff model is too complex to have been widely used for the analysis of experimental results. Dresselhaus and co-workers simplified the Baraff model and succeeded to use it for interpreting their experimental results on magnetoreflection \cite{Dresselhaus1971,Maltz1970,Vecchi1974,Vecchi1976}. Here, we re-introduced the modified Baraff model (called as ``extended Dirac model" in Refs. \cite{ZZhu2011,ZZhu2012,ZZhu2017}), whose energy is given as
\begin{eqnarray}
	E_{n, \sigma}^{\rm e}&=\pm \sqrt{\D^2 + 2\D \ve_{n, \sigma}(k_b)}
	+ \frac{\sigma}{2}g'_b\mu_{\rm B}B,
	\label{modiBaraff}
	\\
	\ve_{n, \sigma}(k_b)&=  \left( n+\frac{1}{2}+ \frac{\sigma}{2} \right)\hbar \omega_c + \frac{\hbar^2 k_b^2}{2m_b},
\end{eqnarray}
where $k_b$ is the wave vector along the magnetic filed and the longitudinal effective mass is
$
	m_b =\bm{b}\cdot \hat{m}\cdot \bm{b}
$
with a unit vector arong the magnetic field $\bm{b}$. The effective mass tensor for electrons at the $L$ point is
\begin{eqnarray}
	\hat{m}=m_0
	\left(
	\begin{array}{ccc}
		m_1 & 0 & 0 \\
		0 & m_2 & m_4 \\
		0 & m_4 & m_3
	\end{array}
	\right).
\end{eqnarray}
The cyclotron frequency and the cyclotron mass is given by
$
	\omega_c =eB/m_c,
$
$
	m_c =\sqrt{{\rm det}\hat{m}/m_b}.
$
The additional g-factor is given as $g_b' = \bm{b}\cdot \hat{g}' \cdot \bm{b}$ and
\begin{eqnarray}
	\hat{g}'&=
	\left(
	\begin{array}{ccc}
		g_1' & 0 & 0 \\
		0 & g_2' & g_4' \\
		0 & g_4' & g_3'
	\end{array}
	\right).
\end{eqnarray}
The first term of Eq. (\ref{modiBaraff}) corresponds to the Cohen-Blount and the Wolff model, which are essentially equivalent to the Dirac electron. (The detailed derivation of this term is given in Ref. \cite{Fuseya2015}.) The second term of the additional g-factor, $g'$, is the contribution from the band other than the two-band.

At high fields, we need a careful treatment for the lowest Landau level (LLL) of $E_{0-}^{\rm e}$. With increasing the magnetic field, the energy gap between the LLLs of the conduction and valence bands can become smaller than the Landau level splittings. In such a high field regime, an interband coupling between two LLLs might be expected \cite{Vecchi1976}. The effect of the inter-LLL coupling can be considered by the form
\begin{eqnarray}
	E_{0-}^{\rm e} &= \pm \sqrt{\left\{ \ve_{0, -} (k_b) - \frac{\tilde{g}'}{2}\mu_{\rm B} B \right\}^2
	+\left(2V\mu_{\rm B}B\right)^2},
	\label{LLL1}
	\\
	\tilde{g}'&=g_b'\left( 1+2\frac{V'|g'|}{\D}\mu_{\rm B}B \right),
	\label{LLL2}
\end{eqnarray}
which are obtained from the model used in Ref. \cite{Vecchi1976}. The parameter $V$ expresses the magnitude of the interband coupling, and $V'$ expresses the correction to $g'$. Both parameters would be given in terms of tensor.

The effective model for hole at the $T$ point is given by the ordinal parabolic dispersion with an effective g-factor in the form
\begin{eqnarray}
	E_0 + \D - E_{n, \sigma}^{\rm h}= \left( n+ \frac{1}{2}\right) \hbar \Omega_c  +
	\frac{\hbar^2 k_b^2}{2M_b}+ \frac{\sigma}{2}G^*_b\mu_{\rm B} B,
	\nonumber\\
\end{eqnarray}
where $E_0$ is the hybridization energy between the conduction band at the $L$ point and the valence band at the $T$ point. The cyclotron frequency $\Omega_c$ and the longitudinal effective mass $M_b$ for holes are given in the same manner as that for electrons. The effective g-factor $G_b^*$ is expressed in terms of the spin mass $M_s$ by
\begin{eqnarray}
	G_b^*=2m_0 \sqrt{\frac{\bm{b}\cdot \hat{M}_s \cdot \bm{b}}{{\rm det}\hat{M}_s}}
\end{eqnarray}
as in Ref. \cite{Smith1964}.
Each model parameter is listed in Table \ref{Table1}. These parameters can give reasonable agreements with experiments up to 65 T for the binary-bisectric plane, and up to 28 T for trigonal-binary and trigonal-bisectrix plane \cite{ZZhu2011,ZZhu2012,ZZhu2017}.

\begin{table}
\caption{\label{Table1} Values of parameters for the effective models  \cite{ZZhu2011,ZZhu2012,ZZhu2017}. For electrons, $m_i$ is the effective mass, $g_i'$ is the additional g-factor. $V_i$ and $V_i'$ are the parameters only for the lowest Landau level of electron. For holes, $M_i$ and $M_{si}$ is the effective orbital and spin mass, respectively. The direct gap at the $L$ point is $2\D$, and the indirect gap between the $L$ and $T$ point is $-E_0$, namely, the hybridization between the electron and hole bands.}
\vspace{2mm}
For electrons:\\
\vspace{-3mm}\\
 \begin{tabular}{ccccc}
 \hline \hline
i & 1 & 2 & 3 & 4 \\
\hline
$m_i$ & 0.00124 & 0.257 & 0.00585 & -0.0277 \\
$g'_i$ & -7.26 & 24.0 & -7.92 & 9.20 \\
$V_i$ & 0.25 & 0.25 & 0.25 & 0.00 \\
$V'_i$ & -0.0688 & -0.0438 & -0.0625 & 0.00 \\
\hline \hline
 \end{tabular}
\\
\vspace{2mm}
\\
\vspace{5mm}
For holes:\\
\vspace{-8mm}\\
 \begin{tabular}{ccc}
 \hline\hline
i & 1 & 3 \\
\hline
$M_i$ & 0.0698 & 0.743  \\
$M_{si}$ & 0.0319 & 10000 \\
\hline \hline
 \end{tabular}
\\
\vspace{2mm}\\
Gap and hybridization:\\
\vspace{-3mm}\\
\begin{tabular}{cc}
\hline
$2\D$ & 15.3 meV
\\
$E_0$ & 38.5 meV
\\
\hline
\end{tabular}
 \end{table}

The computed field dependence of the carrier densities are plotted in Fig. \ref{Fig11}. The carrier density drastically changes keeping the charge neutrality in the magnetic field region near the QL. At around 30 T, it can be more than 5 times larger than the weak field value. This is a characteristic property of compensated semimetals. The different occupation among three electron pockets, the valley polarization, will be discussed later.

\begin{table}
\caption{\label{Table2} Cyclotron and longitudinal mass, and effective g-factor along the principal axes obtained from the angle resolved Landau spectrum up to 12 T. $m_c$ and $M_c$ refer the cyclotron mass of electrons and holes.
$m_b$ and $M_b$ are the band mass of electrons and holes along the field orientation.  They are all normalized by the bare mass of electron, $m_0$.}
\begin{tabular}{lccc}
\\
\hline \hline
  & $\bm{B}\parallel$ Binary &  $\bm{B}\parallel$ Bisectrix &  $\bm{B}\parallel$ Trigonal \\
 \hline
$m_c^{\rm e1}$ & 0.0272 & 0.00189 & 0.0125\\
$m_c^{\rm e2, e3}$ & 0.00218 & 0.00375 & 0.0125\\
$m_b^{\rm e1}$ & 0.00124 & 0.257 & 0.00585\\
$m_b^{\rm e2, e3}$ & 0.193 & 0.0653 & 0.00585\\
$g^{\rm e1}$ & 73.5 & 1060 & 159\\
$g^{\rm e2, e3}$ & 917 & 533 & 159\\
$g'^{\rm e1} $ & -7.26 & 24.0 & -7.92\\
$g'^{\rm e2, e3} $ &16.2 & 0.545 & -7.92\\
\hline
$M_c$ & 0.221 & 0.221 & 0.0678\\
$M_b$ & 0.0678 & 0.0678 & 0.721\\
$G$ & 0.791 & 0.791 & 62.6\\
\end{tabular}
\end{table}

\begin{figure}
	\begin{center}
		\includegraphics[width=7cm]{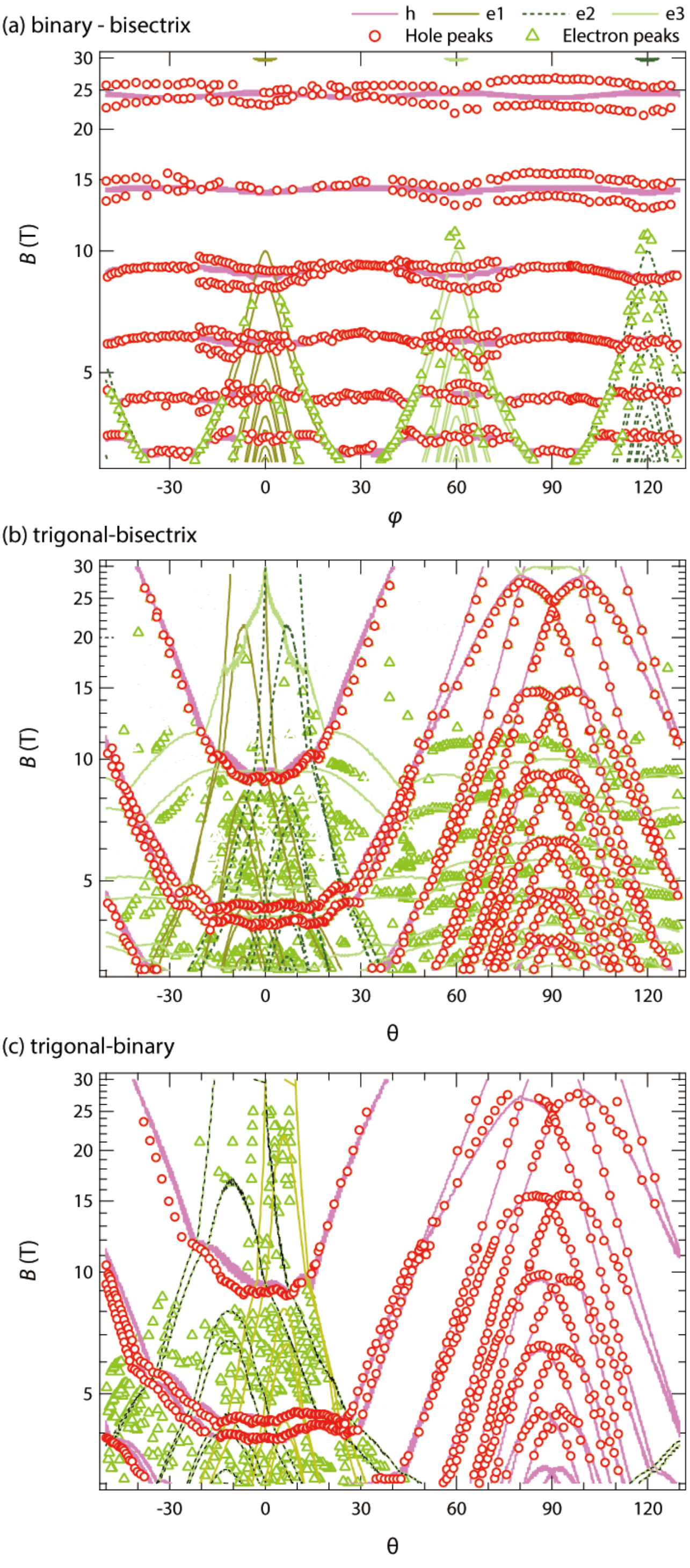}
		\caption{\label{Fig45} Comparison between experimental (symbols) and theoretical (lines) results up to 28T \cite{ZZhu2012}. The additional peaks due to the twinned crystal are not shown in these panels.}
	\end{center}
\end{figure}

Figure \ref{Fig45} shows the comparison of experimental results with theoretical ones for three rotating planes up to 28 T \cite{ZZhu2012}. The additional peaks, which originate from the twinned crystal (discussed in Sec. \ref{Extra}), are discarded in Fig. \ref{Fig45}. (The theoretical results are obtained assuming that $V=0.15$ and $V'$ is isotropic to be -0.0625 \cite{ZZhu2012}.) As seen in Fig. \ref{Fig45}, the agreement is excellent for holes and is less satisfactory for electrons. In the binary-bisectrix plane, the experimentally obtained hole peaks are split because of a small misalignment. In this plane, the cyclotron- and spin-mass for holes are isotropic. Therefore, naively, it is expected that the hole spectrum is independent from the field orientation. However, the Fermi energy is oscillating with respect to the field orientation because of the high anisotropy of electron mass and the charge neutrality, resulting in the slight oscillation also in the hole spectrum.

\subsection{Field-dependent valley population}

When the system attains the QL, all carriers are confined into the LLL, which has the degeneracy of $N_{\rm L}=eB/h$ per unit area. Because of this Landau degeneracy, the carrier density increases as $\propto B^1$ in the QL. If there is only single carrier, $E_{\rm F}$ rapidly decreases in order to keep the carrier number \cite{Owada2017}. On the other hand, in semimetals, electrons and holes can be the reservoir with each other, so that their carrier density increases following the Landau degeneracy. Moreover, when the system has anisotropic multivalleys, the valleys also can be their reservoir with each other. The electrons change their location from one valley to the other valley depending on the orientation of the field. This inter-valley transfer makes the carrier density of each electron valley unbalanced. Like this, in bismuth --- the anisotropic multivalley semimetal --- two kinds of carrier reservoir yield an opportunity of valley polarization.

As one can see in Fig. \ref{Fig11}, the carrier density start to change largely beyond the QL. For $\bB \parallel $ bisectrix, the carrier density of valley e1 increases as $n_{\rm e1}\propto B$ beyond its QL ($B\gtrsim 1.5$ T), and those of e2 and e3 also increases beyond their QL ($B\gtrsim 2.5$ T). The carrier density of e1 is larger than those of e2 and e3 since the LLL of e1 is lower due to the anisotropy of $g'$ ($g'^{\rm e1}=24.0$ and $g'^{\rm e2, e3}=0.545$; cf. Table \ref{Table2}). For $\bB \parallel$ binary, on the other hand, the carrier density of e1 decreases and it remains very low even beyond its QL ($B\gtrsim 10$ T), while those of e2 and e3 increases as $n_{\rm e2, e3}\propto B$ beyond their QL ($B\gtrsim  1.6$ T). This is because the LLL of e1 shifts upward due to the negative sign of $g'$ ($g'^{\rm e1}=-7.26$ and $g'^{\rm e2, e3}=16.2$). Therefore, the anisotropy of $g'$ plays a crucial role for the valley polarization. It should be noted that the valley polarization seen here is different from the valley emptying (100\% valley polarization), where the anisotropy of $V'$ plays a crucial role as will be seen in section \ref{Beyond}.

\subsection{Anisotropic Zeeman splitting due to spin-orbit coupling}
Apart from the valley polarization, another important issue in this field region is the Zeeman splitting of the holes.
What is responsible to the magnitude and anisotropy of the Zeeman splitting or the g-factor is the spin-orbit coupling in crystal. It is shown that the Zeeman splitting is exactly the same as the Landau level splitting, i.e., the ratio of the Zeeman splitting $\D E_Z$ to the cyclotron energy $\hbar \omega_c$, dubbed $M\equiv \D E_Z/\hbar \omega_c$ is $M=1$ for every orientation of magnetic field according to the two-band $\kp$ theory, which is equivalent to the Dirac model (Sec. \ref{Effective model}) \cite{Cohen1960,Wolff1964,Fuseya2015}. There is no reason for $M$ to be larger than unity as far as the theory is based on the two-band model. Actually, if one looks at the value of $M$ for the electrons at the $L$ point in bismuth, the observed value of $M$ is very close to unity and isotropic as is shown in Table \ref{table3}.

For holes at the $T$ point in bismuth, on the other hand, the properties of $M$ cannot been explained by the previous theory for g-factor. First, $M$ is extremely anisotropic, and second, it largely exceeds unity in one configuration (Table \ref{table3}). These two puzzles have been experimentally confirmed by numerous studies since half a century ago \cite{Smith1964,Edelman1976,Bompadre2001,Behnia2007_PRL,ZZhu2011,ZZhu2012}.
\begin{table}
\caption{\label{table3} Values of $M\equiv \Delta E_Z /\hbar \omega_c$ evaluated with the cyclotron mass and the g-factor shown in Table \ref{Table2}.}
\begin{tabular}{lccc}
\\
\hline \hline
  & $\bm{B}\parallel$ Binary &  $\bm{B}\parallel$ Bisectrix &  $\bm{B}\parallel$ Trigonal \\
 \hline
$M^{\rm e1}$ & 0.90 & 1.02 &  0.950\\
$M^{\rm e2, e3}$ & 1.01 &  1.00 &  0.950\\
\hline
$M^{\rm h}$ & 0.0875 & 0.0875 & 2.12
\end{tabular}
\end{table}
This longstanding puzzle was finally solved by using the general formula for the g-factor newly obtained based on relativistic multiband $\kp$ theory \cite{Fuseya2015}. It gives not only a qualitative interpretation to the large and anisotropic g-factor, but also a quantitative agreement with experimental results by combining the tight-binding model of Liu and Allen \cite{Liu1995}.

The anisotropy of the g-factor is basically determined by the property of the matrix elements of velocity operator, such as $\langle \psi_{i \sigma}| \bm{v} |\psi_{j \sigma'}\rangle$. ($i, j$ are indices of Bloch bands.) By taking into account the symmetry property of the $T$ point in bismuth, it is shown that the g-factor is finite for $\bB \parallel$ trigonal, while it is exactly zero for $\bB \perp$ trigonal, which explains the anisotropy of $M$.

The large magnitude of $M$ is explained by taking into account the contribution from many bands. If one take into account only two bands, $M$ is always unity. On the other hand, it was shown that it can be larger and smaller than unity if one take into account more than two bands \cite{Fuseya2015b}. Surprisingly, in the case of hole at the $T$ point, a band 1 eV far from the valence band can enhance the magnitude of $M$ by a factor of 2. This significant effect is a resultant of the interband effect of spin-orbit coupling.
The theoretical value by a combination of the relativistic multiband $\kp$ theory and the tight-binding model of Liu and Allen \cite{Liu1995} is $M=2.08$, which agrees quite well with experimental value of $M=2.12$. Not only for the pure bismuth and at ambient pressure, but also for Sb substitution and under a pressure, the theory can give results agrees with experiments \cite{Fuseya2015b}.

\subsection{``Extra" peaks in Landau spectrum}\label{Extra}
There is another longstanding mystery for bismuth near the QL. Several decades ago, Mase and co-works found some extra peaks in the ultrasonic attenuation spectrum \cite{Sakai1969,Matsumoto1970,Mase1971,Mase1980}. The extra peaks were observed in the field as large as 10 T off from the trigonal axis, and they could not be assigned to any known Landau levels. Following this observation, a possibility of the field-induced excitonic insulator had been discussed by many authors \cite{Maki1971,Fukuyama1971b,Fukuyama1971c,Kajimura1975,Yoshioka1978,Kuramoto1979}, but a clear evidence for the phase transition had not been obtained after all.

Since 2007, there has been a renewal of interest in the Landau spectrum of bismuth at higher magnetic fields above 10 T \cite{Behnia2007_Science}. Another extra peak structures were found in the Nernst response at high magnetic fields exceeding the QL \cite{Behnia2007_Science,HYang2010}. Generally speaking, no Landau spectrum is expected at fields beyond the QL. Therefore, the extra peaks in the Nernst response cannot be attributed to any Landau level. This anomaly was also found in the magneto-torque measurement \cite{LLi2008}. Note that the first report of torque magnetometry study claimed that the anomaly accompanies a hysteresis suggesting the first-order \cite{LLi2008}, while subsequent study of torque magnetometry \cite{Fauque2009} and magnetostriction \cite{Kuchler2014} failed to reproduce the hysteresis.

Several theories has been proposed for this problem of extra peaks. Sharlai and Mikitik proposed that the extra peaks are due to the misalignment of the field \cite{Sharlai2009}. But this possibility was ruled out by the angular dependence of the Nernst response \cite{ZZhu2011}. Seradjeh, Wu and Phillips speculated that the extra peaks can be the signal of surface states \cite{Seradjeh2009}. But this was also ruled out since the amplitude and angular dependence of the Nernst signal totally exhibit the bulk nature \cite{Behnia2010,ZZhu2011}. Alicea and Balents argued the possible instabilities toward the charge-density-wave and Wigner crystal phase, and suggested the hysteresis found in Ref. \cite{LLi2008} should be originated from the latter phase \cite{Alicea2009}. However, as was mentioned, such a phase transition has not been reproduced so far except for the first report of torque magnetometry.

\begin{figure}
	\begin{center}
		\includegraphics[width=7cm]{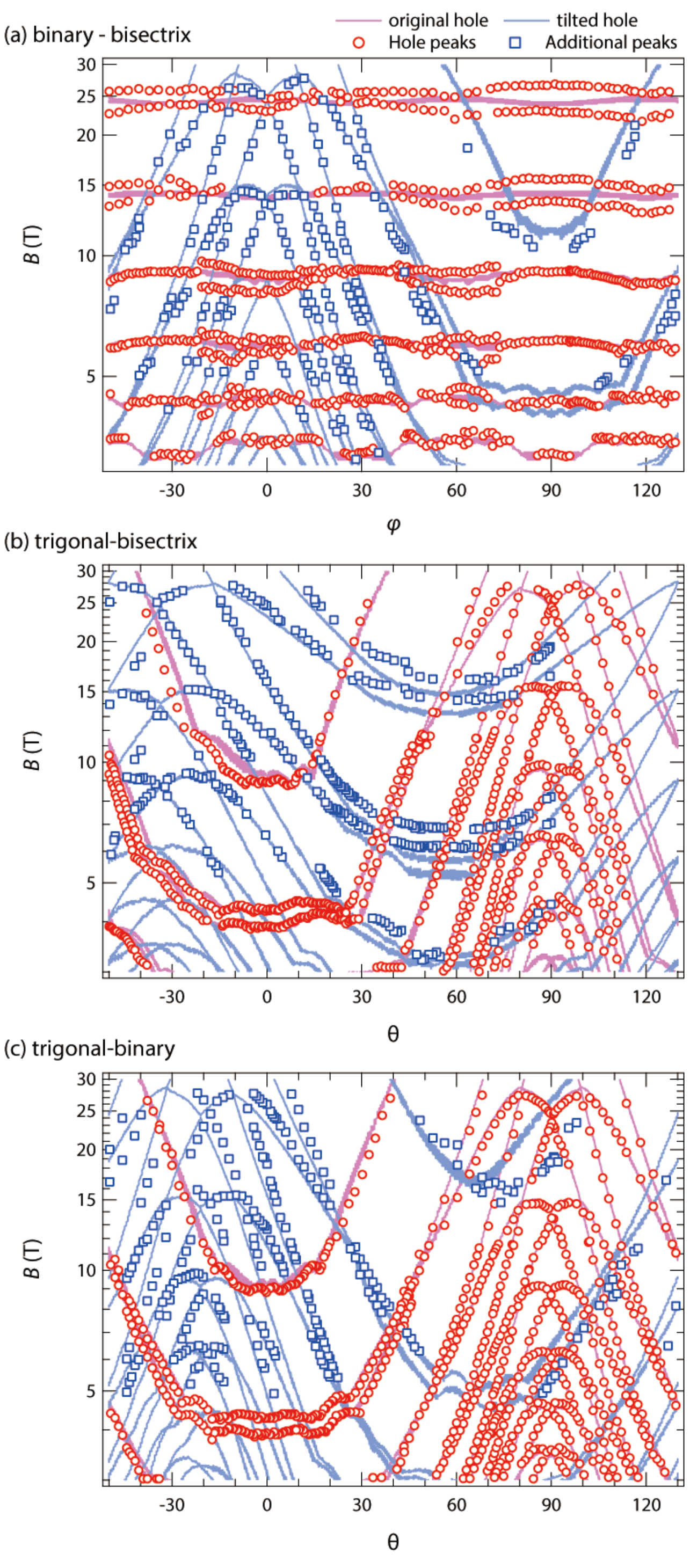}
		\caption{\label{Fig49} Experimental Landau spectrum for holes and the extra peaks (symbols), and theoretical spectrum (solid lines) for holes in the original and twinned crystal in the three rotating planes.}
	\end{center}
\end{figure}

This problem was eventually resolved by extending the map of angle resolved Landau spectrum for entire solid angle up to 28 T, and by taking into account contributions from the secondary twinned crystal \cite{ZZhu2012}.
When there are two domains in a crystal, there should be a secondary contributions to the signals in addition to the primary signal from the main crystal. Such a twin boundary has been actually detected by scanning tunneling microscopy measurement on bismuth \cite{Edelman1996}.

In Fig. \ref{Fig49}, the peak positions for primary hole (circles) and the extra peaks (squares) obtained from the Nernst response are shown. The solid lines are the theoretical spectrum for holes (the same ones shown in Fig. \ref{Fig45}) and the hole spectrum from the secondary crystal tilted $108.4^\circ$ from the primary crystal. As seen in Fig. \ref{Fig49}, the agreement is perfect, which guarantees the validity of the twinning scenario. It should be commented here that the extra peaks found in the ultrasonic attenuation measurements nearly correspond to the extra peaks detected by the Nernst response, so that they are most probably caused by twins. This is also true of the anomaly at 40 T observed in the transverse magnetoresistance \cite{Fauque2009}.
Most probably, this is also the case of anomalies seen in the magnetoresistance of Bi$_{96}$Sb$_{4}$ beyond its quantum limit \cite{Banerjee2008}.
Consequently, one-particle non-interacting  theory can successfully reproduce the angle resolved Landau spectrum obtained by various experiments in the high field region near the QL \cite{ZZhu2011,ZZhu2012,Kuchler2014}, including those previously attributed to electron interaction \cite{Behnia2007_PRL,LLi2008,HYang2010}.  Note also that the hysteresis observed in torque magnetometry experiments reported in Ref. \cite{LLi2008} and attributed to a first-order phase transition was not reproduced by subsequent experiments \cite{Fauque2009,Kuchler2014}.

\section{Tuning contribution of valleys III: Beyond the quantum limit}\label{Beyond}
The dramatic occurrence of integer and fractional Hall effect at high fields in the two-dimensional electron gas leads to a naive question: What is the fate of the three-dimensional electron gas far beyond the QL? This has been studied both theoretically and experimentally, yet only a few aspects has been understood. So far, possibility of spin density wave \cite{Celli1965}, charge density wave \cite{Fukuyama1978,Yoshioka1981,Heinonen1986,MacDonald1987}, Wigner crystal \cite{Kleppmann1975,Kuramoto1978,Schlottmann1979}, valley density wave \cite{Tesanovic1987}, excitonic insulators \cite{Fenton1968,Abrikosov1970a,Abrikosov1970b,Fukuyama1971b,Fukuyama1971c} have been discussed theoretically. Experimentally, on the other hand, graphite is the only three-dimensional system in which electronic instabilities have been detected \cite{Tanuma1981,Iye1982,Yaguchi2009,Fauque2013b,Akiba2015,Behnia2015,ZZhu2017b,Arnold2017,LeBoeuf2017}.

\subsection{Sudden drop of magnetoresistance}

The magnetoresistance measurements on bismuth in the field region far beyond the QL (up to 90.5 T for $\bB \parallel$ binary and up to 65 T for field rotation in the binary-bisecrix plane) indicate that there is no such an electronic instability occurs in bismuth \cite{ZZhu2017}. Instead, they reveal that one or two valleys become totally empty \cite{ZZhu2017}.
A possibility of controlling the valley polarization was initially discussed in two-dimensional electrons in AlAs quantum wells \cite{Shkolnikov2002,Shayegan2006}, and subsequently in various systems \cite{DXiao2007,Isberg2013,XXu2014,Jo2014,Renard2015}. However, none of these can totally dry up the Fermi sea. It is only in bismuth that electrons attain 100\% valley polarization.

\begin{figure}
	\begin{center}
		\includegraphics[width=9.5cm]{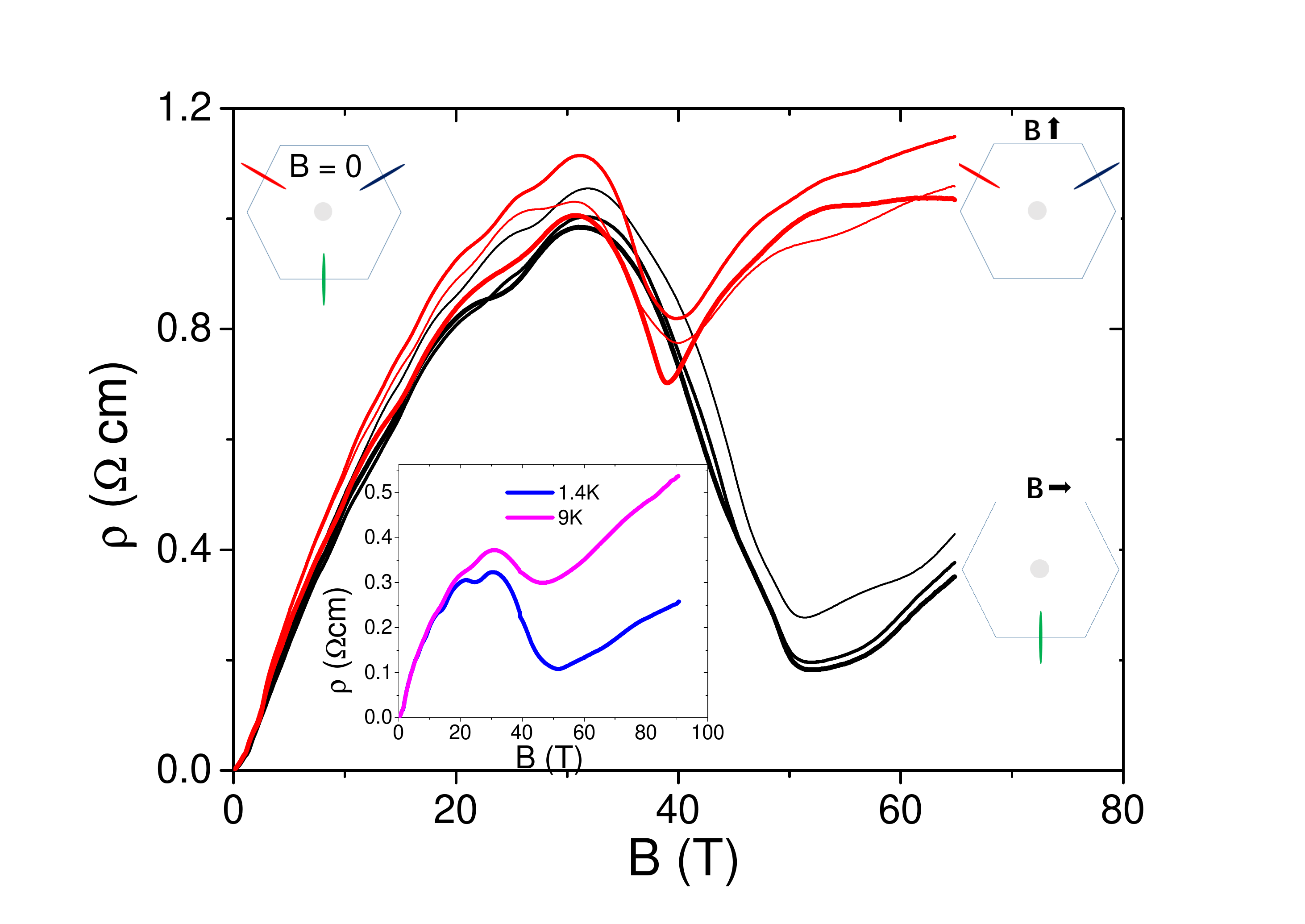}
		\caption{\label{Fig51} Transverse magnetoresistance of a bismuth crystal up to 65 T at $T=1.56$ K for magnetic fields along the three equivalent binary (black) and bisectrix (red) crystalline axes. The illustrations indicates the situation after the emptying valley. For the field along binary (bisectrix), two (one) valleys are emptied and one (two) valley remains. The inset shows the magnetoresistance up to 90.5 T with a field along binary and a current along bisectrix. The system remains metallic in the whole field range.}
	\end{center}
\end{figure}
Figure \ref{Fig51} shows the transverse magnetoresistance up to 65 T at $T=1.55$ K for magnetic fields along the three equivalent binary ($\varphi=0, \pm 2\pi/3$, black) and bisectrix ($\varphi=\pi/2, -\pi/6, -5\pi/6$, red) axes with $\bm{j}\parallel$ trigonal axis. The inset of Fig. \ref{Fig51} is the magnetoresistance up to 90.5 T for $\bB \parallel $ binary and $\bm{j} \parallel$ bisectrix at 1.4 K (magenta) and 9 K (blue). The magnetoresistance keeps increasing with magnetic field up to 35 T both for $\bB \parallel$ binary and bisectrix axes.

The magnetoresistace stop to increase and begins to drop at 35 T. The drop is very anisotropic. The magnetoresistance for $\bB \parallel$ binary keeps to drop until 50 T and the amplitude is 4 times larger than that for $\bB \parallel$ bisectrix, where the magnetoresistance keeps to drop until 40 T.
The onset of this drop was observed in many years ago \cite{Hiruma1983,Miura1994}. It has been concluded that the drop at 35 T is attributed to the quantum oscillation due to the hole.
However, the measurement at higher fields and $360^\circ$ angle reveals that the idea of hole oscillation is unlikely to be valid. The large difference between the binary and bisectrix, the large amplitude, and the large frequency of the oscillation cannot be interpreted at all. Moreover, the absence of the metal-insulator transition up to 90.5 T (inset of Fig. \ref{Fig51}) contradicts with their hole scenario, which predicts the metal-insulator transition below 90 T for $\bB\parallel$ binary axis \cite{Hiruma1983}.

\subsection{Angle resolved Landau spectrum}
In order to resolve the mystery of the sudden drop in magnetoresistance, the theoretical model was fine tuned for such a high magnetic field. In this field region far beyond the QL, the property of the LLL is crucial.
For the clarification of the property of the LLL, it would be quite helpful to analyze the angle-resolved Landau spectrum as has been done for the other Landau levels.

\begin{figure}
	\begin{center}
		\includegraphics[width=9.5cm]{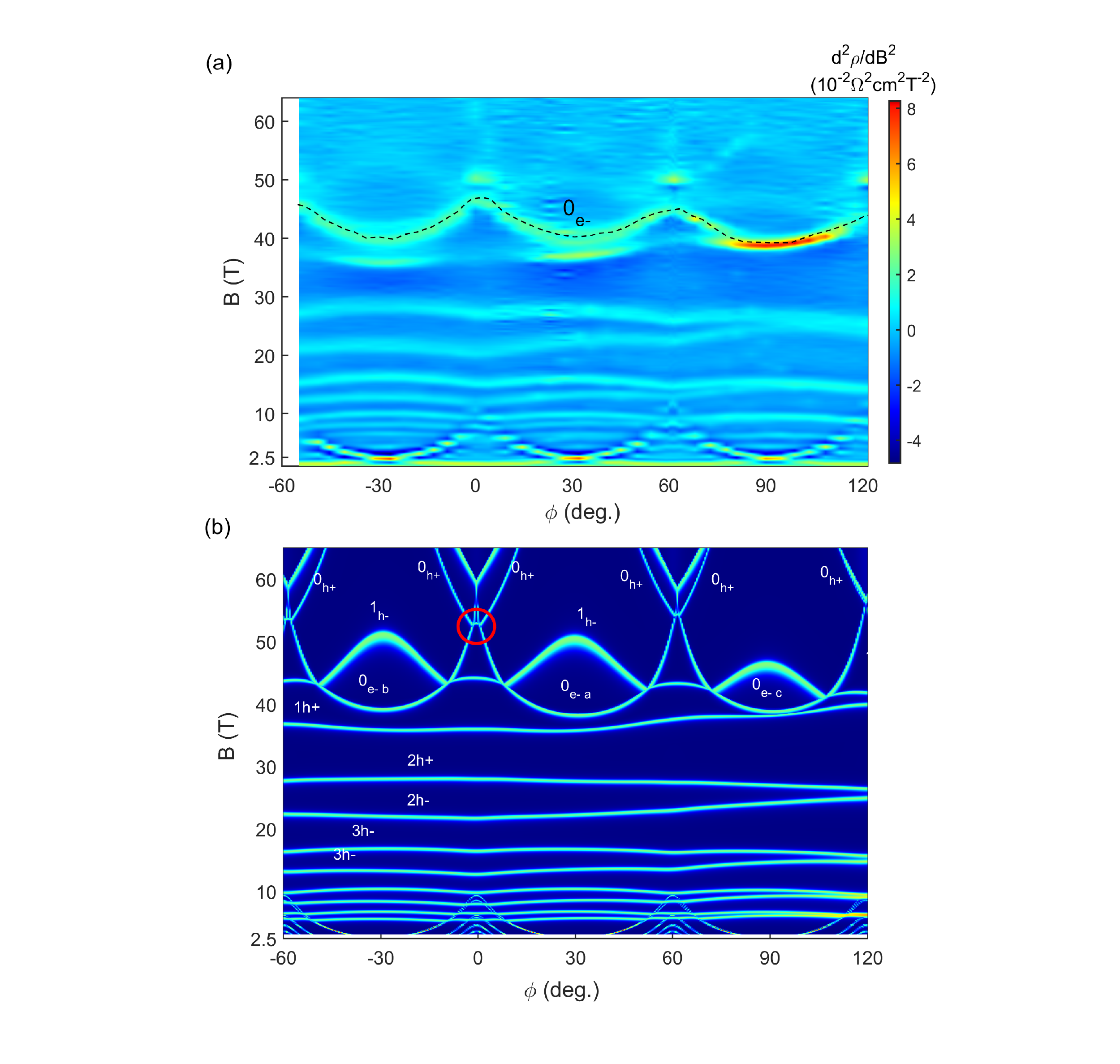}
		\caption{\label{Fig53} Comparison of experimtal and theoretical Landau spectrum up to 65 T.
		(a) Color plot of the second derivative of magnetoresistance, $d^2 \rho / dB^2$. The bright lines corresponds to the holes/electrons Landau levels intersecting the Fermi level. The $0_{\rm e-}$ is marked with a black dashed line.
		(b) Color plot of the theoretical Landau spectrum assuming a small misalignment. Some of bright lines of Landau spectrum are indexed. A valley is emptied after $0_{\rm e-}$ crosses the Fermi level. Three Landau levels (two electron- and one hole-pocket) simultaneously cross the Fermi level at the region marked with a red circle.}
	\end{center}
\end{figure}

The color map of the angular dependence of the second derivative of the magnetoresistance is plotted in Fig. \ref{Fig53} (a), which corresponds to the angle-resolved Landau spectrum \cite{ZZhu2017}. The Landau spectrum obtained by the magnetoresistance agrees with that obtained by the Nernst response below 28 T (Fig. \ref{Fig45}) \cite{ZZhu2012}.
The Zeeman splitting of the hole spectrum is due to the slight misalignment, while it should vanish when the field is perfectly aligned in the binary-bisectrix plane \cite{Fuseya2015b}.  The experimentally obtained angle-resolved Landau spectrum is to be compared with theoretically obtained one shown in Fig. \ref{Fig53} (b). The excellent agreement validates the parameters used in the effective model listed in Table \ref{Table1}. Note that the misalignment of the magnetic field is also taken into account in Fig. \ref{Fig53} (b).

\subsection{Emptying a valley (100\% valley polarization)}

\begin{figure}
	\begin{center}
		\includegraphics[width=8.5cm]{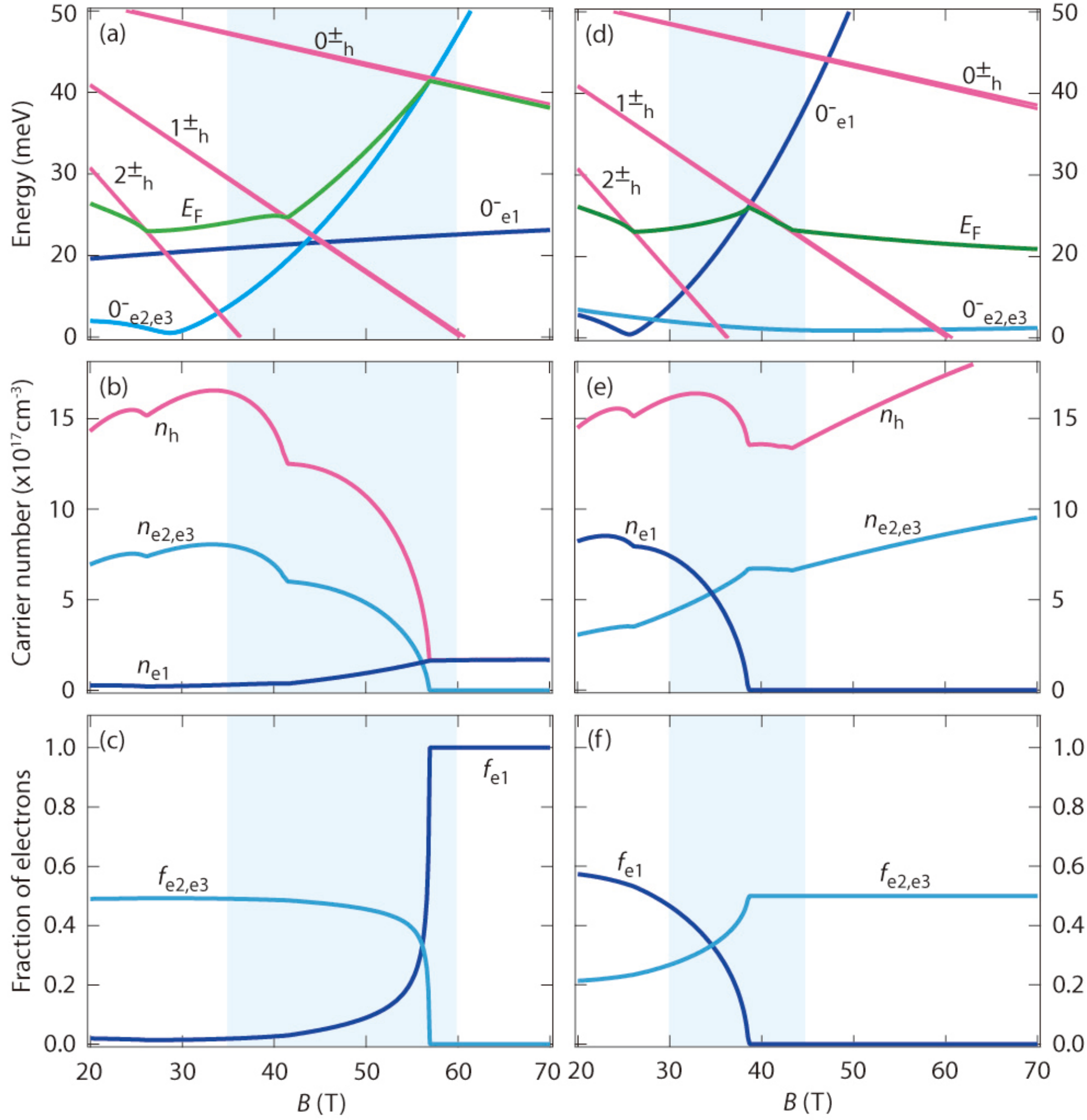}
		\caption{\label{Fig54} Magnetic field dependence of (a), (d) the Landau levels and the Fermi energy; (b), (e) the carrier density; and (c), (f) the proportion of carriers in different electron pockets for $\bB \parallel$ binary and $\bB \parallel$ bisectrix. The shaded region corresponds to the magnetic field region, where the experimental magnetoresistance drops. In these calculations, the field is assumed to be perfectly oriented to $\bB \perp$ trigonal.}
	\end{center}
\end{figure}

\begin{figure}
	\begin{center}
		\includegraphics[width=7cm]{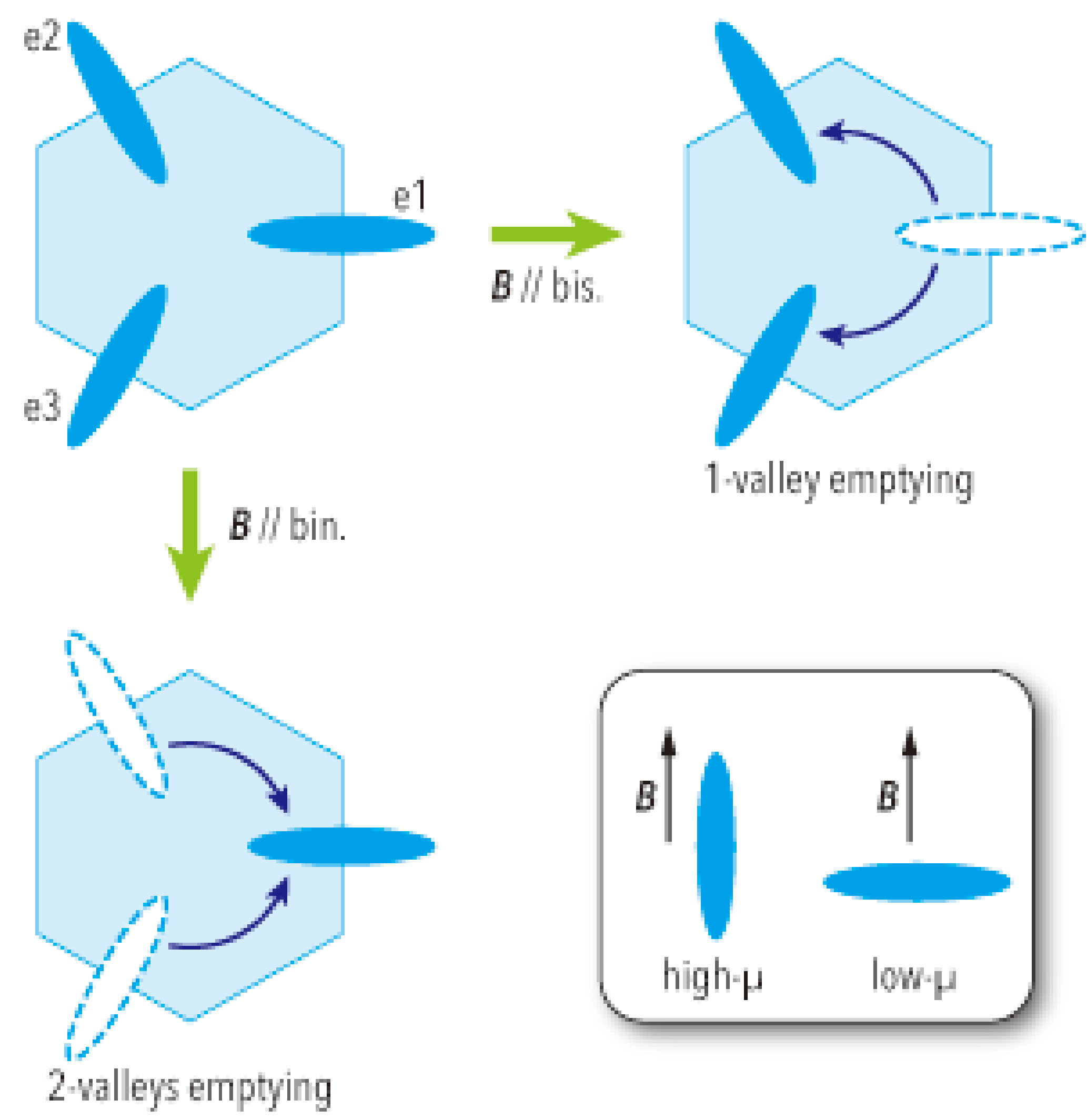}
		\caption{\label{Fig55} Illustrations of the emptying valleys of bismuth with different orientation of the applying filed. When the high magnetic field ($\sim $ 40-50 T) is applied along the binary (bisectrix) axis, two (one) valleys are emptied. By this emptying, the electrons in the high mobility valleys are transfered to the low mobility valley. Note that the relevant mobility of the anisotropic valley depends on the field orientation. The relevant mobility is higher when the filed is along its major axis.}
	\end{center}
\end{figure}

The theoretical Landau levels and the Fermi energy are plotted as a function of the magnetic field in Fig. \ref{Fig54} (a) for $\bB \parallel$ binary and (d) for $\bB \parallel$ bisectrix. Whole electron carriers are confined into their LLL above 12 T of $\bB \perp$ trigonal. 
For $B \gtrsim 10$ T, the quadratic correction of the additional g-factor, the $V'$-term in Eq. (\ref{LLL2}), dominates the behavior of the LLL (especially when the field is perpendicular to the longer axis of the electron ellipsoid). The LLL shifts upward rapidly with increasing the magnetic field. This upturn of the LLL is observed more directly by the magnetoreflection measurement \cite{Vecchi1976}, though the upturn is more rapid than that is shown in Fig. \ref{Fig54} (a). The Fermi energy largely shift upward accompanyed by the upturn of the electron LLL.
The evolution of the Fermi energy results in the drastic change in the carrier density as shown in Fig. \ref{Fig54} (b) and (e). The hole carrier density (i.e., the total electron carrier density) increases up to 35 T by a factor of 5, then suddenly decreases.

At 55 T of $\bB \parallel$ binary, the Fermi energy finally crosses two LLL, so that the two valleys are emptied, namely, 100\% valley polarization is achieved. On the other hand, for $\bB \parallel $ bisectrix, one valley is emptied at 40 T. The distribution of carriers among the electron pockets is shown in Fig \ref{Fig54} (c) and (f). For $\bB \parallel$ binary, and below 35 T, most of the electron carriers locates in valleys e2 and e3, but there still remains carrier in valley e1. Then, the carriers in valley e2 and e3 move into e1, and the whole electron carriers are confined into the valley e1 above 55 T. For $\bB \parallel$ bisectirix, on the other hand, the electron carriers in valley e1 move into valleys e2 and e3 above 40 T. These situations are illustrated in Fig. \ref{Fig55}. We can control the two-valley emptying and one-valley emptying only by changing the orientation of the magnetic field.

The shaded region in Fig. \ref{Fig54} corresponds to the magnetic field window where the experimental magnetoresistance drops. Figure \ref{Fig54} indicates that, in this colored region, the carriers abruptly changes their accommodation from one valley to the other. Then, why the move of electrons makes the magnetoresistance drop? The origin of the drop can be understood intuitively as follows.

At high magnetic fields, the magnetoresistivity is proportional to the mobility perpendicular to the field and the current direction [cf. Eq. (\ref{MC1})]. When the field is parallel to the longer axis of the ellipsoid, the relevant component of the mobility tensor is high, while it is low when the field is perpendicular to the longer axis as is shown in the box of Fig. \ref{Fig55}. For example,
when $\bB \parallel $ binary, the mobility of valley e2, e3 are high, and that of e1 is low. Through the electron move from e2, e3 to e1, the relevant mobility tensor changes from higher to lower, resulting in the drop of the magnetoresistance. When $\bB \parallel$ bisectrix, the mobility of valley e1 is high and that of e2, e3 is low. Also in this case, the electron move from e1 to e2, e3 causes the magnetoresistance drop. However, there is a difference between the two field orientations: For $\bB \parallel $ binary, nearly 100\% ($=50\% + 50\%$) electrons lose their high mobility, while only about 60\% electrons lose their high mobility for $\bB \parallel$ bisectrix. This is why the magnetoresistance drop is more drastic for $\bB \parallel $ binary than bisectrix. Therefore, the initial puzzles, the large difference between the binary and bisectrix, the large amplitude and the large width of the drop, has been totally interpreted by taking into account the 100\% valley polarization and the highly anisotropic mobilities. (A more detailed derivation for the magnetoresistance drop is given in supplementary information of Ref. \cite{ZZhu2017}.)

We should note the similarity between the present effect and the so-called Gunn effect \cite{Gunn1963}. In some III-V semiconductors such as GaAs and InP, the resistance decreases after an ``electric" field reaches a threshold level. The basic mechanism of this Gunn effect is understood as the transfer of electrons from high mobility valleys to low mobility valleys by the electric field \cite{Butcher1967}. The effect discussed here is the transfer of electrons between valleys with different mobilities caused by magnetic (not electric) field.

\subsection{Field dependence of mobility}
A quantitative explanation of the mangetoresistance drop requires an accurate knowledge of the field dependence of the mobility. The magnetic-filed dependence of the mobility has been a longstanding problem, and it attracts renewed interests recent days in relation to the linear magnetoresistance \cite{Abrikosov1969,Abrikosov2003,Song2015}. In the case of the magnetoresistance of bismuth, the constant relaxation time approximation (i.e., the constant mobility) is not enough to give a quantitative interpretation. Here we examine how the semiclassical framework can give quantitative interpretations by considering the field dependence of the mobility.

\begin{figure}
	\begin{center}
		\includegraphics[width=7cm]{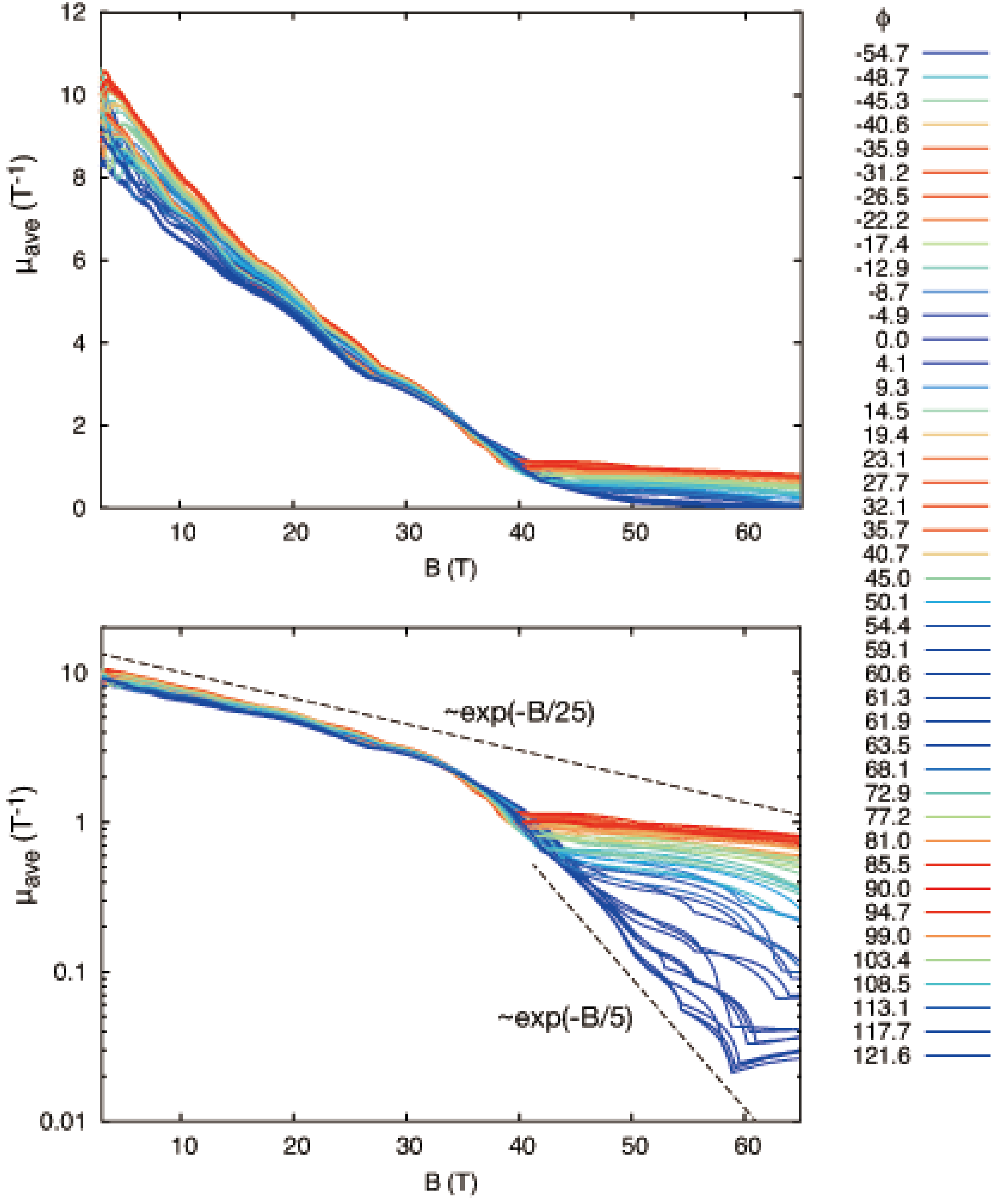}
		\caption{\label{Fig56} Averaged mobility $\mu_{\rm ave}$ as a function of magnetic field for different orientations in a linear (upper panel) and semi-logarithmic (lower panel) scale. $\phi=0$ corresponds to the filed along the bisectrix axis.}
	\end{center}
\end{figure}

The field dependence of the ``averaged" mobility can be estimated from the relation
\begin{eqnarray}
	\rho (B) =\frac{\mu_{\rm ave}(B) B^2}{en_h},
\end{eqnarray}
by combining the experimental result of $\rho(B)$ and the theoretical result of $n_h (B)$. Figure \ref{Fig56} shows the field dependence of $\mu_{\rm ave}(B)$ so obtained for each orientation of magnetic field. The field dependence for every orientation follows $\mu_{\rm ave}\sim e^{- B/\beta_0}$ below 30 T and much faster after wards. ($\beta_0$ is a constant parameter.) The main reason for this change around 30 T is that the relative weight of the components of the mobility tensor changes as the valley begins to be emptied.

\begin{figure}
	\begin{center}
		\includegraphics[width=7cm]{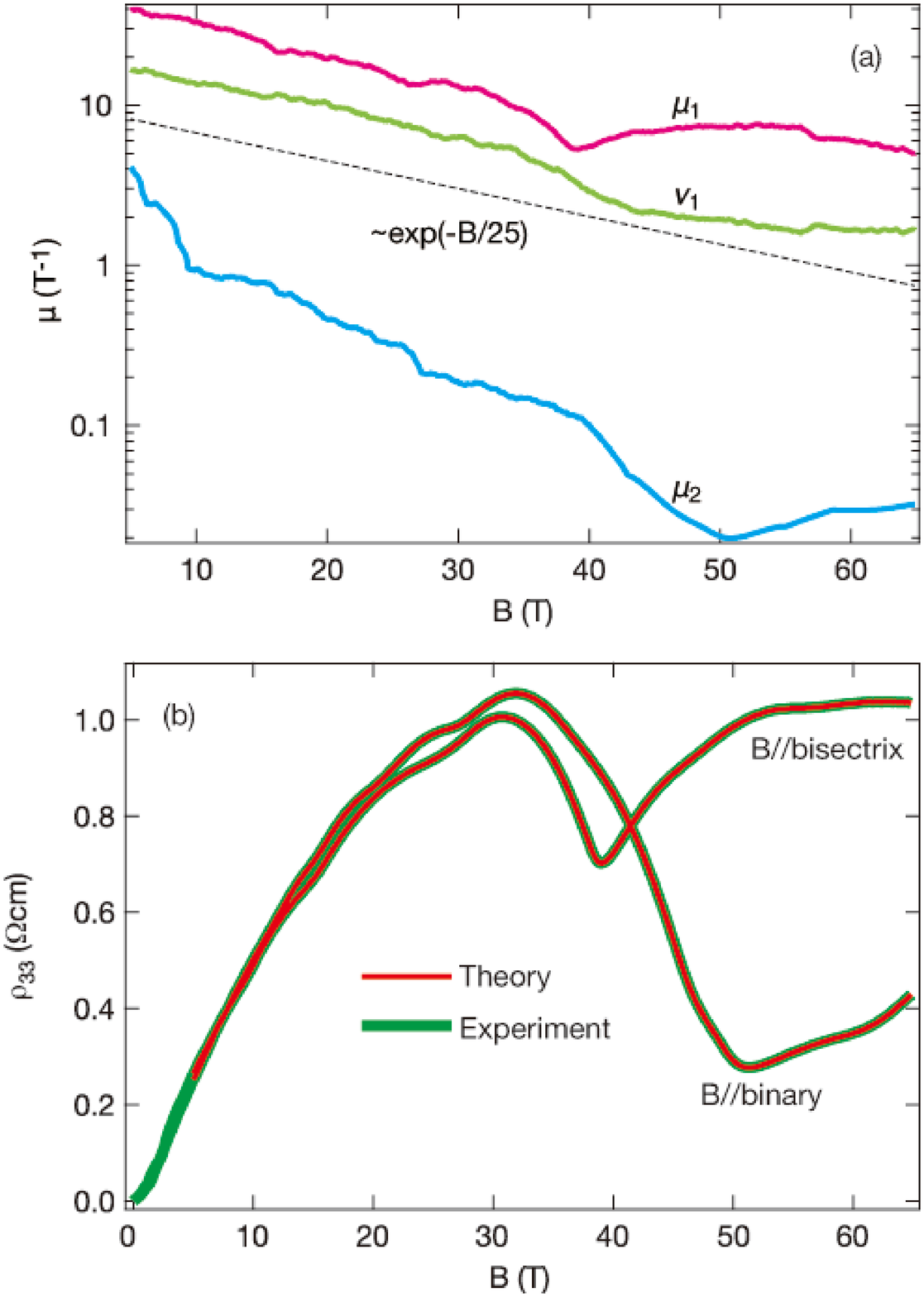}
		\caption{\label{Fig57} (a) Field dependence of each mobility ($\mu_1$, $\mu_2$, and $\nu_1$) determined so as to fit the experimental magnetoresistance. (b) Magnetoresistance so obtained theoretically (thin red lines) and experimentally (thick green lines). A single set of mobilities can give the theoretical magnetoresistance which agrees perfectly with experiments for both orientation of the magnetic field. }
	\end{center}
\end{figure}

Of course, the mobility of bismuth is highly anisotropic, so that the field dependence for each component of the mobility should be taken into account independently. Figure \ref{Fig57} (a) shows the field dependence of each mobilities determined so as to fit the experimental magnetoresistivity using $n_{\rm e1}$ and $n_{\rm e2, e3}$ obtained theoretically.
(The detailed calculations for the individual field dependence of mobility is given in supplementary information of Ref. \cite{ZZhu2017}.) It also exhibits roughly $\mu_i \sim e^{-B/\beta_0}$ dependence, which can give a theoretical result perfectly agree with experimental magnetoresistance.

To summarize, the semiclassical theory combined with a phenomenological assumption on the field dependence of the mobility tensor can give theoretical results close to the experimental ones. The microscopic origin of the field dependence of the mobility is an open question.

\section{Spontaneous valley symmetry breaking}
So far, we have looked at the recent progress on the magnetoresistance and the quantum oscillations on bismuth in three different magnetic field regions, and seen that the most of the properties can be interpreted based on the one-particle picture of semiclassical theory and the Landau levels obtained by relativistic multiband $\kp$ theory. Here we review on a new phenomena which has not been explained theoretically --- the valley symmetry breaking. It should be distinguished from the valley polarization in the following sense. In the valley polarization, the population of valley is different among equivalent valleys for one orientation of magnetic field. However, the physical quantities between crystallographycally equivalent orientations (there are three equivalent binary and bisectrix axes in bismuth) is the same in the valley polarized state. In the valley symmetry broken state, on the other hand, the physical quantities among crystallographycally equivalent orientations are different, namely, the system breaks the underlying crystal symmetry. In this section, we shall review the experimental signatures for this valley symmetry broken state.

\subsection{Transport signatures}
The experimental configuration is the same as the situation of Fig. \ref{Fig31} (a), i.e., the electric current flows parallel to the trigonal axis, and the magnetic field is rotated in the binary-bisectrix plane. 
As is argued in Sec. \ref{Low field limit}, the angular dependence of the magnetoresistance exhibit a sixfold symmetry in a wide range of temperature at low fields, reflecting the $C_{\rm 3v}$ symmetry of the crystal [Fig. \ref{Fig31} (a)]. However, at low temperatures and under relatively high magnetic fields, this symmetry is found to be lost \cite{ZZhu2011b,Collaudin2015}.

\begin{figure}
	\begin{center}
		\includegraphics[width=8cm]{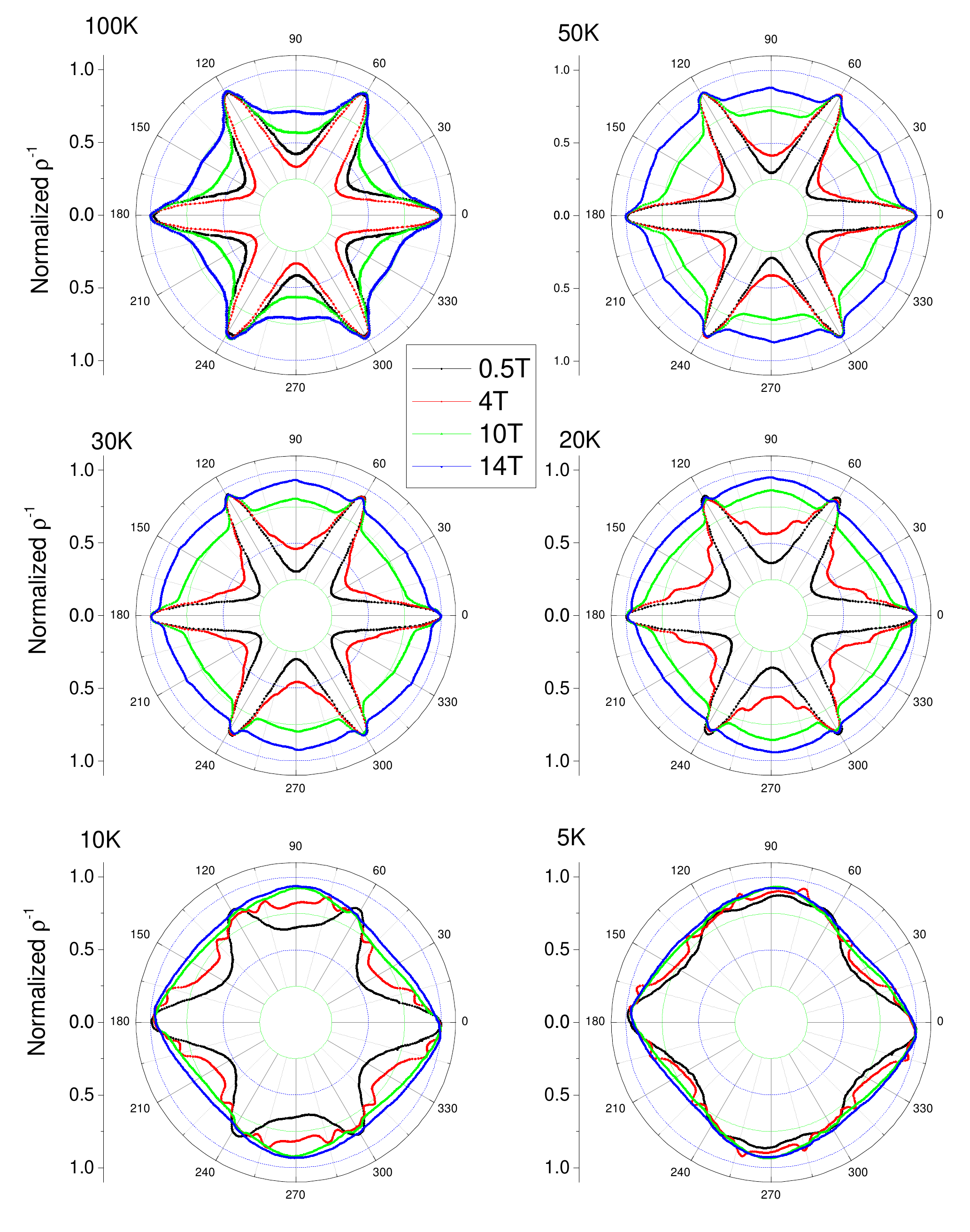}
		\caption{\label{Fig61} Evolution of the angle-dependent magnetoresistance ($\rho^{-1}$) with decreasing temperature. Each panel shows a polar plot of the inves of magnetoresistance normalized to its maximum value at a given magnetic field. At high temperature, the threefold symmetry of the underlying lattice is preserved. As the temperature decreases, this symmetry is lost above a threshold magnetic field, which decreases in amplitude with cooling.}
	\end{center}
\end{figure}

Figure \ref{Fig61} shows the polar plots of the inverse of magnetoresistance, $\rho^{-1}$, at different temperatures and magnetic fields. The results are normalized to the maximum value of $\rho^{-1}$ at each temperature in order to make the comparison easier. At high temperatures ($T>30$ K), the sixfold symmetry is clearly preserved for all magnetic fields, but it begins to lost around 10 K. The system completely lost the sixfold symmetry at 5 K for all magnetic field, though the twofold symmetry remains. Thus, exactly speaking, what is lost is the threefold symmetry.

\begin{figure}
	\begin{center}
		\includegraphics[width=8cm]{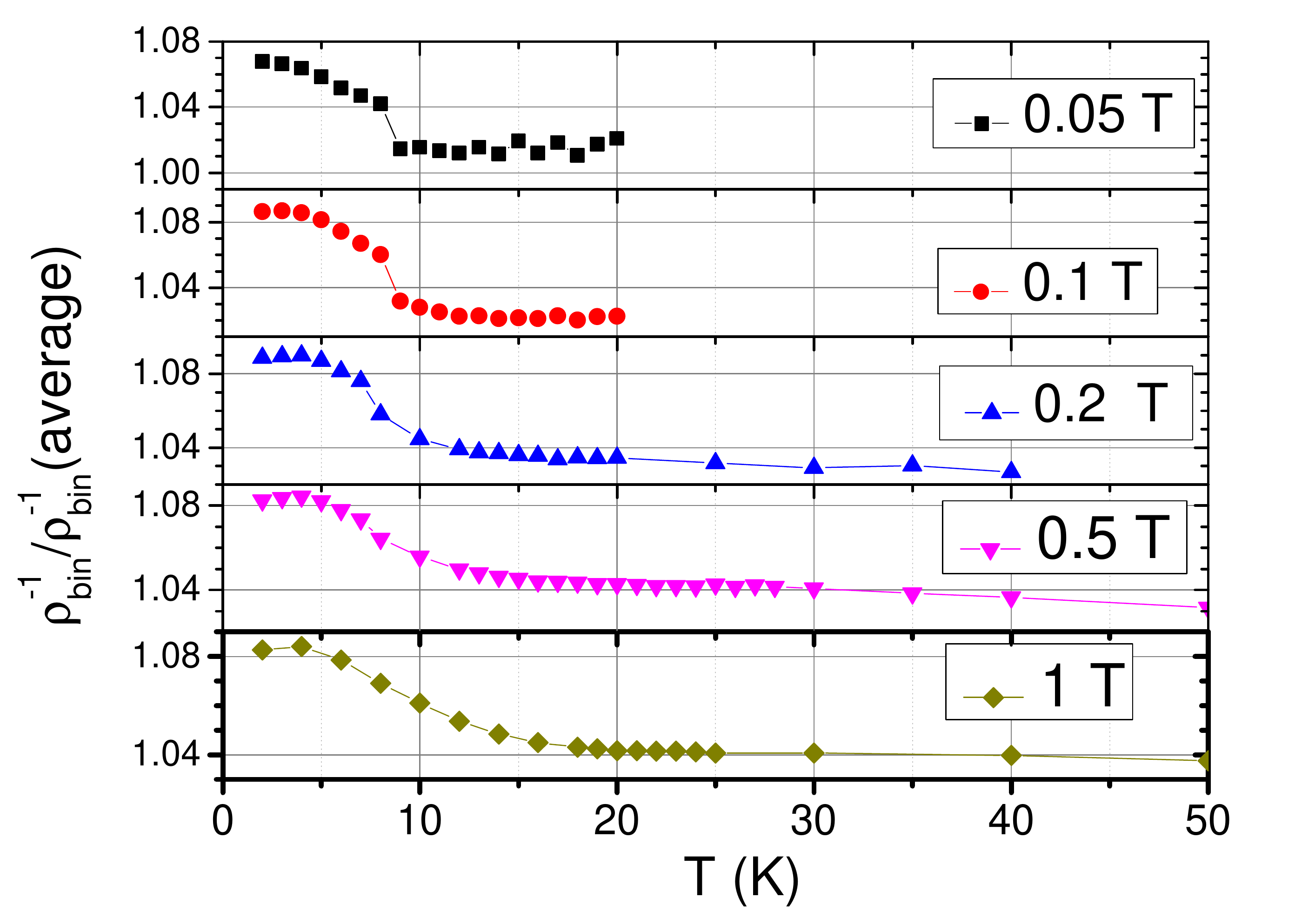}
		\caption{\label{Fig64} Ratio of $\rho_{\rm bin1}^{-1}$ to $(\rho_{\rm bin1}^{-1}+\rho_{\rm bin2}^{-1}+\rho_{\rm bin2}^{-1})/3$ as a function of temperature. A clear jump in the ratio is observed at low fields, suggesting the phase transition. The ``transition" becomes wider and shifts to higher temperature as the magnetic field increases. }
	\end{center}
\end{figure}
\begin{figure}
	\begin{center}
		\includegraphics[width=8cm]{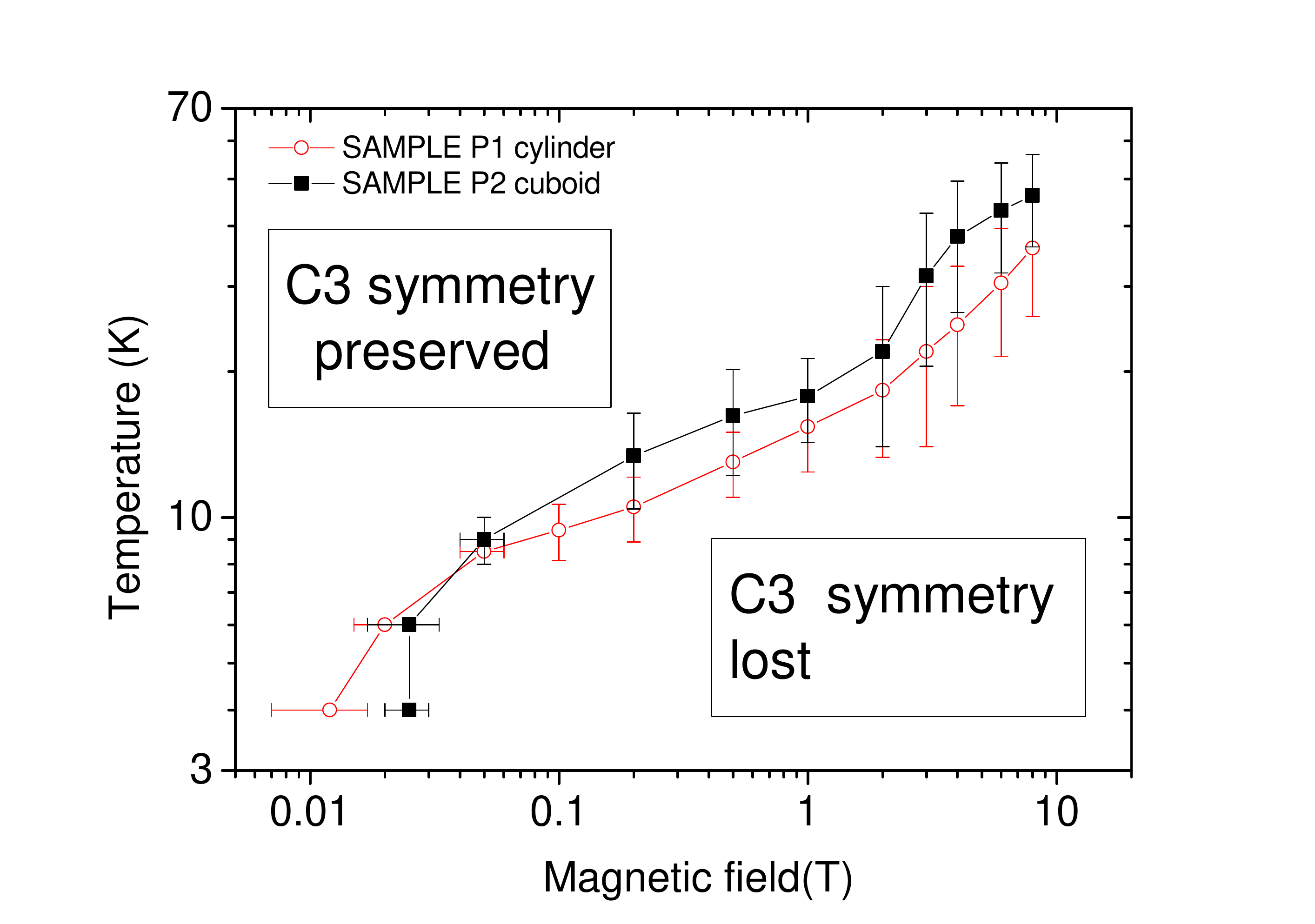}
		\caption{\label{Fig65} Variation of the ``transition" temperature for the loss of threefold symmetry as a function of the magnetic field in the two samples with different shapes: cylinder and cuboid. The $C_3$ symmetry is preserved in the low-field and high-temperature region and lost in the high-filed and low-temperature region.}
	\end{center}
\end{figure}

The evolution of this loss of symmetry becomes more visible when one plots the relative magnitude of magnetoresistance for the field aligned along one binary axis. Figure \ref{Fig64} shows the temperature dependence of the ratio of $\rho_{\rm bin1}^{-1}$ to the average value among three equivalent binary axes, $(\rho_{\rm bin1}^{-1}+\rho_{\rm bin2}^{-1}+\rho_{\rm bin2}^{-1})/3$. At high temperatures, the ratio is almost constant. However, it suddenly deviates from the average value at a certain threshold temperature. At low fields, $B\lesssim 1$ T, there is a jump in the ratio at low temperature $T\lesssim 10 $ K, strongly suggests a phase transition. With increasing the magnetic field, the ``transition" temperature shifts to higher temperatures and it becomes wider. At a field as high as 8 T, the deviation in the ratio occurs over a temperature range as wide as 20 K. On the other hand, the maximum value of the ratio in the low temperature limit is independent from the magnitude of the field. It should be emphasized here that no evidence for hysteresis has been ever found.
From these data, the transition-like temperature is plotted as a function of magnetic field in Fig. \ref{Fig65}. Each symbol represents the intersection between the low-temperature and high-temperature behaviors. In the low-temperature and high-field region, the $C_3$ symmetry is lost, while, in the high-temperature and low-filed region, it is preserved.

By investigating the quantum oscillation of resistivity (the Shubnikov-de Haas effect), it has been clarified that the frequency of oscillation is identical among the three equivalent orientations, even though there is a difference in the amplitude of oscillation. This is consistent with the results by the Nernst oscillations \cite{HYang2010} and the magnetostriction argued below.

\subsection{Thermodynamic signatures}

The loss of threefold symmetry in bismuth was first reported in the magnetoresistance measurements \cite{ZZhu2011b}. Soon later, it was also observed by the magnetostriction measurements, which gives the thermodynamic evidence of the loss of threefold symmetry state \cite{Kuchler2014}.
By tilting the field angle from the trigonal axis ($\theta=0$), the peak position in the field dependence of the magnetostriction coefficient, $-\lambda (B)$, is symmetric between $\theta >0$ and $\theta <0$. However, the hight of the peak is not symmetric between $\theta >0$ and $\theta <0$. The detailed analysis of the magnetostriction spectrum reveals that, at each peak, the contribution from one valley (e3) is larger than the other (e1).

The magnetostriction coefficient, $-\lambda$, is directly couples to the differential density change $ -\partial \Delta N/\partial B$. Therefore, the different peak height indicates the difference in the density of states between three equivalent valleys. This is the first thermodynamic evidence of the loss of threefold symmetry.

\subsection{In search of a theory}

The valley symmetry broken state discussing here is characterized by the obserbation that the amplitude of the physical quantities, such as the resistivity and the magnetostriction coefficient, loses the symmetry of the underlying crystal.
The valley symmetry breaking observed in bismuth further shows the following features, which are basically common in both transport and thermodynamical measurements.
First, the symmetry breaking appears only in the low-temperature and high-field region. The temperature dependence in resistivity shows an abrupt change reminiscent of a phase transition. Second, the size of the crystallographycally equivalent valleys are equal, i.e., there is no valley polarization with respect to the carrier density. Third, there is no trace of hysteresis through the loss of threefold symmetry, although there is a jump in the deviation of physical quantity from the average value of equivalent valleys.

The possibility that the loss of threefold symmetry is due to the misalignment has been ruled out in one carries out by the two-axis rotation experiment \cite{ZZhu2011}. The possibility of the uncontrolled strain, which is responsible for lifting the valley degeneracy, can be also ruled out by the facts that the frequency of quantum oscillation by resistivity and magnetostriction remains identical among the crystallographycally equivalent valleys.

Then, what is the origin of the valley symmetry breaking? So far, while several theoretical ideas have been proposed, none of them gives a satisfactory explanation of the whole spectrum of the observations.

One possibility is to follow the scenario of valley nematicity proposed by Abanin {\it et al.} for  quantum Hall systems with multivalleys \cite{Abanin2010}. According to their theory, anisotropy of the dispersion relation favors the state where all carriers are concentrated into one valley. For that nematic valley ordering, the anisotropy of the mass of each valley and the Coulomb interaction play important roles. Although their theory is calculated for two-dimensional systems, in the last part of their paper, they mentioned that their idea could apply to the three dimensional case of bismuth. However, their nematic valley ordered state should exhibit valley polarization, where the carrier densities of three valleys are unequal. This is a serious disagreement with the breaking of the valley-symmetry observed in bismuth.

K\"uchler {\it et al.} \cite{Kuchler2014} recalled the idea that the present situation is reminiscent of disordered semiconductors, where the Coulomb interaction in a disordered system can open a gap very close to the chemical potential \cite{Efros1975}. It is still an open question if this idea is valid also for the multivalley system such as bismuth.

Another possibility would be the idea of the field-induced lattice distortion due to electron-phonon coupling proposed by Mikitik and Sharlai \cite{Mikitik2015}. They showed that a first-order phase transition accompanying a spontaneous symmetry breaking of magnetostriction can occur when the Landau level touches the Fermi energy due to the electron-phonon interaction. However, the field-induced lattice distortion and a series of the structural transition, which are crucial consequence of their theory, has not been observed until now.

\section{Conclusions}
The main conclusion of this review consists of two parts. First, from the experimental point of view, the angular dependence of magnetoresistance in wide range of magnetic field is essential to resolve the electronic state under a magnetic field, especially for the system with multivalley. The measurements applying fields along the high symmetric axes cannot solve the valley degeneracy, which prevents us from a unique determination of the Landau levels. Even the data with misalignment is sometimes helpful to resolve the Landau spectrum.
Second, from the theoretical point of view, the conventional one-particle picture, semiclassical theory and $\kp$ theory under the field, can give quantitative agreements with experiments unexpectedly perfectly. This is true even when the experimental data seems to be unexplainable at first glance. One can learn from the long history of studies on bismuth that it is not too late to conclude the new data indicates unconventional phenomena after an elaborate analysis based on the conventional one-particle theory.

Bismuth exhibits three different magneto-properties according to the region of magnetic field.
In the low field limit, bismuth exhibits remarkable angular dependences (even at room temperature) \cite{ZZhu2011,Collaudin2015}, which is interpreted by the semiclassical theory almost perfectly \cite{Aubrey1971,Collaudin2015}. The angular oscillation of magnetoresistance can change its shape by changing temperature because of the different temperature dependences of mobilities.
Near the QL, high anisotropy of energy dispersions causes the (partial) valley polarization, which is also interpreted by the Landau spectrum obtained by the relativistic multiband $\kp$ theory with fine parameter tunings \cite{ZZhu2011,ZZhu2012,Kuchler2014}. There has been two kinds of longstanding mysteries in this field region. The large and anisotropic Zeeman splitting for holes \cite{Smith1964,Edelman1976,Bompadre2001,Behnia2007_PRL,ZZhu2011,ZZhu2012} is explained as the effect of large interband spin-orbit coupling \cite{Fuseya2015b}. The extra peaks in Landau spectrum \cite{Sakai1969,Matsumoto1970,Mase1971,Mase1980,Behnia2007_Science,HYang2010,LLi2008} is understood as the signal from the twinned crystals \cite{ZZhu2012}.
Beyond the QL, the valley emptying (100\% valley polarization) occurs due to the anisotropic LLL motions. This is the first observation of the complete valley-emptying by a magnetic field \cite{ZZhu2017}. One can easily control one- or two-valley emptying only by changing the orientation of the magnetic field. The valley emptying causes the abrupt drop in the magnetoresistance, which is also due to the high anisotropy of the mobility.

Besides these successful findings and understandings, a further problem on magnetoresistance is the field dependence of mobility. Although there are some proposals for the microscopic origin of the quasi-linear magnetoresistance \cite{Abrikosov1969,Abrikosov2003,Song2015}, no satisfactory theory can explain quantitatively the field dependence of magnetoresistance on bismuth for a wide region of magnetic field.

A new finding in bismuth is the valley symmetry breaking (loss of three fold symmetry) at low temperatures and at moderately high magnetic field \cite{ZZhu2011b,Collaudin2015}. Although some possible scenarios have been proposed \cite{Abanin2010,Mikitik2015}, the loss of three fold symmetry without both the hysteresis and the valley polarization has not been explained yet. The search for the mechanism of the valley symmetry breaking has only just begun.

\section*{Acknowledgments}
We would like to thank A Banerjee, A Collaudin, H Fukuyama, J P Issi, W Kang, B Lenoir, R D McDonald, M Ogata, and H Yang.
for their precious collaboration during the experimental and theoretical investigation on bismuth in the last decades.
This work was part of SUPERFIELD and QUANTUM LIMIT projects funded by Agence Nationale de la Recherche.
ZZ is supported by the 1000 Youth Talents Plan, the National Science Foundation of China (Grant No. 11574097), the National Key Research and Development Program of China (Grant No.2016YFA0401704), and by directors' funding grant number 20120772 at LANL.
BF acknowledges support from Jeunes Equipes de l'Institut de Physique du College de France (JEIP).
KB was supported by China High- end foreign expert program, 111 Program and Fonds-ESPCI-Paris.
YF was supported by JSPS KAKENHI grants 16K05437, 15KK0155 and 15H02108.

\section*{References}

\bibliographystyle{iopart-num}
\bibliography{Bismuth}

\end{document}